# The APM Galaxy Survey. IV: Redshifts of Rich Clusters of Galaxies

G. B. Dalton, G. Efstathiou, S. J. Maddox & W. J. Sutherland*
*Department of Physics, University of Oxford, Keble Road, Oxford, OX1 3RH. UK.*

**ABSTRACT**
We present redshifts for a sample of 229 clusters selected from the APM Galaxy Survey, 189 of which are new redshift determinations. Non-cluster galaxy redshifts have been rejected from this sample using a likelihood ratio test based on the projected and apparent magnitude distributions of the cluster fields. We test this technique using cluster fields in which redshifts have been measured for more than 10 galaxies. Our redshift sample is nearly complete and has been used in previous papers to study the three dimensional distribution of rich clusters of galaxies. 157 of the clusters in our sample are listed in the Abell catalogue or supplement, and the remainder are new cluster identifications.

## 1 INTRODUCTION

Clusters of galaxies can provide much information on the large-scale distribution of matter in the universe. The cluster catalogues of Abell (1958), Zwicky et al. (1968) and Abell, Corwin & Olowin (1989) were constructed by visual inspection of sky suvey plates, and have been used widely to investigate cluster properties, galaxy evolution, and the large-scale structure of the universe.

Studies of the three-dimensional distribution of clusters in the Abell catalogues have shown that it is anisotropic, with an excess number of cluster pairs found with small angular separations but large radial separations (Sutherland 1988; Sutherland & Efstathiou 1991; Efstathiou et al. 1992). This effect may be an artefact either of the process of selecting galaxy clusters from inhomogeneous plate material, of the enhancement of a cluster due to the overlap of adjacent cluster halos (Dekel et al. 1989), of the process of selecting clusters by eye over long periods, or of patchy Galactic extinction or contamination by 'clusters' which are simply projected associations along the line of sight (Peacock & West 1992). This effect produces more power on large scales than is predicted by models where structure forms by purely gravitational mechanisms, given the observed amplitude of the galaxy autocorrelation function (Maddox, Efstathiou & Sutherland 1993)

We have constructed cluster catalogues from the APM Galaxy Survey (Maddox et al. (1990a,b)) with the aim of im-proving on Abell's catalogues and resolving the issue of the large-scale clustering of clusters. In this paper we list new measurements of redshifts for galaxies in 189 clusters, and a catalogue of 229 cluster redshifts. This catalogue encompases the cluster sample used for the determination of the correlation amplitude for the APM clusters (Dalton et al. 1992; Efstathiou et al. 1992). In §2 we give a brief description of the cluster selection algorithm. In §3 we describe the maximum likelihood estimator for the cluster distance which we use to test our redshift observations for the presence of foreground galaxies. The observations are discussed in §5 and we summarise our observations and discuss further extension of the survey in §6.

## 2 THE CLUSTER CATLOGUE

The APM Galaxy Survey (Maddox et al. (1990a,b)) is based on scans of 185 UK Schmidt J plates with the SERC Automatic Plate Measuring (APM) machine and lists accurate positions and magnitudes for over 2 million galaxies brighter than a magnitude limit of $b_J = 20.5$, with completeness $\sim 90$–$95\%$, stellar contamination $\sim 5\%$ and negligible dependence of the galaxy surface density on declination or galactic latitude. The survey covers a solid angle of 4300 square degrees.

Clusters of galaxies were selected from the APM Survey using a two-stage process (a detailed description will be given in paper V in this series). The first stage involves locating dense spots in the galaxy surface density above a magnitude limit $b_J = 20.5$ by applying a percolation algorithm linking together galaxies with angular separations less than 0.7 times the mean separation. This algorithm was applied separately to each photometrically matched Schmidt plate, and groups containing $\geq 20$ galaxies were chosen as candidate cluster centers.

* Current Address: Center for Particle Astrophysics, University of California, Berkeley, Ca 94720, USA

In the second stage, an iterative procedure was applied to define a characteristic magnitude $m_X$ and richness $\mathcal{R}$ for each cluster as described by Dalton *et al.* (1992). The cluster selection algorithm is objective, and improves on Abell's selection in in several ways. Our smaller counting radius (half the Abell radius) and the weighting scheme improve the contrast above background and reduces the overlap area of neighbouring cluster halos and the associated ambiguity of assigning galaxies to clusters in supercluster regions.. Using our characteristic magnitude, $m_X$, instead of Abell's $m_{10}$ means that the distance estimate should be nearly independent of cluster richness. The high photometric accuracy of the APM Survey should yield $m_X$ estimates and richnesses that are uniform over the entire area of the survey.

## 3 CLUSTER DISTANCES USING MAXIMUM LIKELIHOOD

### 3.1 Cluster Model

In order to arrive at an estimator for the cluster distance we first adopt a model for the galaxy distribution within the cluster field. We adopt a Schechter function for the galaxy luminosity function, with parameters determined by Loveday *et al.* (1992):

$$\phi(L)dL = \phi_*(L/L_*)^\alpha \exp(-L/L_*)dL/L_*, \quad (1)$$

$B_{J*} = -19.6; \quad \phi_* = 1.2 \times 10^{-2} h^3 \text{Mpc}^{-3}; \quad \alpha = -1.08.$

We will discuss the choice of these parameters in §5. The mean galaxy density at redshift $z$ is then given by

$$\begin{aligned} \bar{n} &= \phi_* \int_{x_{lim}(z)}^{\infty} x^\alpha \exp(-x) dx, \\ &= \phi_* \Gamma(\alpha+1, x_{lim}(z)), \\ x_{lim}(z) &= L_{min}(z, m_{lim})/L_*, \end{aligned} \quad (2)$$

where $L_{min}(z,m)$ is the minimum luminosity visible at a redshift $z$, given an apparent magnitude limit of $m$.

We assume that the luminosity of a cluster galaxy is independent of position within the cluster, and so our cluster model may be expressed in terms of seperate functions of apparent magnitude and radial position. As an estimate of a mean cluster radial profile we take the form of the cluster-galaxy cross-correlation given by Lilje & Efstathiou (1988)

$$\xi_{cg}(r) = (r/r_0)^{-\varepsilon}; \quad \varepsilon = 2.2, r_0 = 8.8 \ h^{-1} \text{Mpc}, \quad (3)$$

where, for the purposes of the model, we will assume that $\xi_{cg} \equiv 0$ beyond a limiting radius $r_T$.

The surface density of galaxies around the cluster centre may be expressed as

$$\begin{aligned} \mu(\sigma) &= 2 \int_0^\infty n(r) dy \\ r^2 &= \sigma^2 + y^2, \end{aligned} \quad (4)$$

where $\sigma$ and $y$ are the components of the radial vector on the sky and along the line of sight, respectively.

Inserting (3) into (4) and assuming $\xi_{cg} \gg 1$ gives

$$\mu(\sigma) = \frac{\bar{n} r_0^\varepsilon}{\sigma^{\varepsilon-1}} \int_0^{r_T^2/\sigma^2 - 1} \frac{t^{-1/2} dt}{(1+t)^{\varepsilon/2}}. \quad (5)$$

In all cases of interest we have $\sigma \ll r_T$, and so

$$\mu(\sigma) \approx \frac{\bar{n} r_0^\varepsilon}{\sigma^{\varepsilon-1}} \int_0^\infty \frac{t^{-1/2} dt}{(1+t)^{\varepsilon/2}}. \quad (6)$$

The number of galaxies associated with the cluster is then

$$N_c(\sigma) = 2\pi^2 r_0^\varepsilon \phi_* \Gamma(\alpha+1, x_{lim}(z)) \frac{\Gamma((\varepsilon-1)/2)}{\Gamma(\varepsilon/2)} \frac{\sigma^{3-\varepsilon}}{(3-\varepsilon)}. \quad (7)$$

Note that in this equation we have assumed that the galaxy luminosity function is the same for galaxies in clusters and in the field. We shall return to this question later.

### 3.2 The Maximum Likelihood Estimator

From equation 7 the total number of galaxies belonging to a cluster at redshift $z$ predicted to lie within a projected radius $\theta$ to a limiting magnitude $m$ is

$$N_c(\theta, m) = A(z)\Gamma(\alpha+1, x_{min}(z,m))\theta^{3-\varepsilon}, \quad (8)$$

and from the field,

$$N_f(\theta, m) = B(z)\theta^2. \quad (9)$$

Here we have defined

$$\begin{aligned} A(z) &= \frac{2\pi^2 \phi_* r_0^\varepsilon}{3-\varepsilon} \frac{\Gamma((\varepsilon-1)/2)}{\Gamma(\varepsilon/2)} (d_A(z))^{3-\varepsilon}, & (10a) \\ d_A(z) &= y(z)/(1+z), & (10b) \\ B(z) &= \pi \phi_* \int_0^\infty \int_{x_{min}(z,m)}^\infty x^\alpha e^{-x} dx y^2 dy, & (10c) \\ x_{min}(z,m) &= 10^{0.4(m_*(z)-m)}, & (10d) \\ m_*(z) &= M_* + 25 + 5\log(d_l(z)) + K_z z, & (10e) \\ d_l(z) &= (1+z)y(z), & (10f) \end{aligned}$$

where we adopt a K-correction term of $K_z = 3.0$. Differentiating equations 8 and 9 gives

$$\frac{dN_c(\theta,m)}{d\theta dm} = 0.4 \ln(10) A(z)(3-\varepsilon)\theta^{(2-\varepsilon)}[x_m^\alpha \exp(-x_m)] \quad (11)$$

$$\frac{dN_f(\theta,m)}{d\theta dm} = 0.8 \ln(10) \pi \phi_* \theta \int_0^\infty x_m^{\alpha+1} \exp(-x_m) y^2 dy, \quad (12)$$

so that for each galaxy found within a circle of radius $\theta_f$ we can define the quantity

$$p_i = N_g \frac{dN_c(\theta_i, m_i) + dN_f(\theta_i, m_i)}{N_c(\theta_f, m_{lim}) + N_f(\theta_f, m_{lim})}, \quad (13)$$

where $N_g$ is the total number of galaxies found within $\theta_f$, and hence we may define the likelihood of the field representing a cluster at redshift $z$ by

$$\mathcal{L} = \prod_i p_i, \quad (14)$$

and obtain an estimate of the cluster redshift by maximising $\mathcal{L}(z)$.

Equation (13) assumes that the field galaxy distribution is unclustered, and that all clusters have the same richness. To modify the first of these assumptions we obtain the field normalisation, $\alpha$, by determining the ratio of the local galaxy background surface density around the cluster to that predicted for a uniform distribution with our input luminosity function. The normalisation of the cluster model is then fixed by the total number of galaxies in the field to be

$$\beta = \frac{N_g - \alpha N_f(\theta_f, m_{lim})}{N_c(\theta_f, m_{lim})}, \quad (15)$$

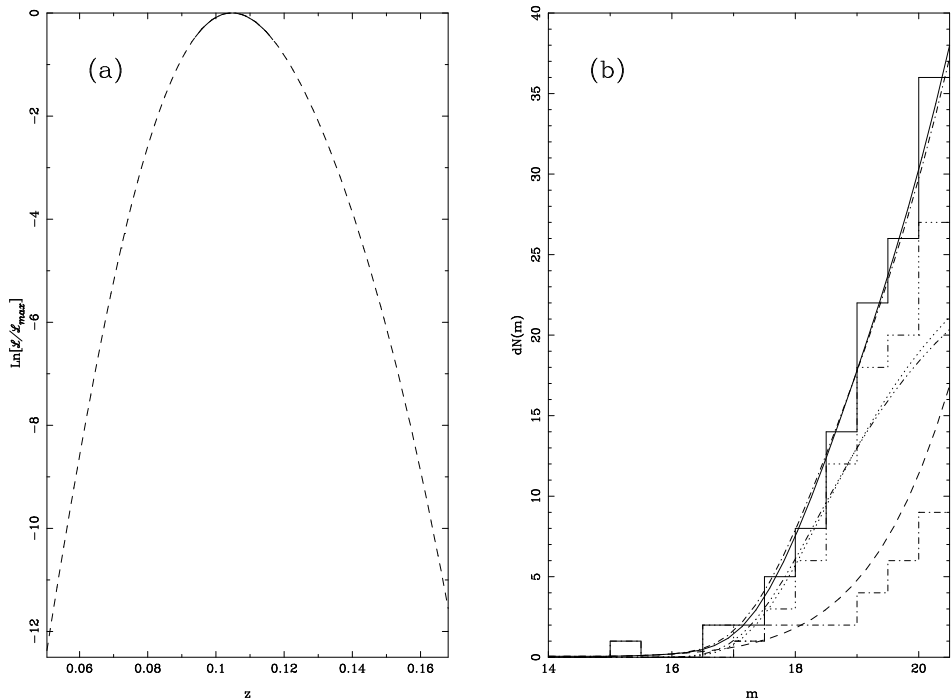

**Figure 1.** a) The likelihood function for a model cluster generated with $z = 0.1$. The solid part of the curve represents a 1-$\sigma$ interval as described in the text. The likelihood function is maximised for $z = 0.1042$. b) The apparent magnitude distribution of the field of the cluster (solid histogram), made up of cluster (dot-dot-dot-dashed) and field (dot-dashed) components. The solid line shows the apparent magnitude distribution for the maximum likelihood redshift, with cluster and field components shown (dotted and dashed lines, respectively). The dot-dashed and dot-dot-dot-dashed lines show the model predictions for the total apparent magnitude distribution and cluster component given $z = 0.1$, given the adopted background.

which may be used as an alternative estimate of the cluster richness. We can now rewrite (13) as

$$p_i = N_g \frac{\beta dN_c(\theta_i, m_i) + \alpha dN_f(\theta_i, m_i)}{\beta N_c(\theta_f, m_{lim}) + \alpha N_f(\theta_f, m_{lim})}. \quad (16)$$

### 3.3 Some Examples

We generated a set of model clusters using the model described above to give an estimate of the accuracy of the redshift estimator, and to provide comparison data for real cluster fields. The likelihood function for one model cluster is shown in Fig.1a. The estimated redshift agrees well with the true redshift for the cluster. In Fig.1b we show the apparent magnitude distribution of the cluster and field components compared to those predicted for the redshift of maximum likelihood. Since, in the limit of large numbers, the quantity $-2\ln \mathcal{L}/\mathcal{L}_{max}$ is expected to approximate $\chi^2$ distributed with 1 degree of freedom we obtain a 1-$\sigma$ confidence interval by computing the range of redshifts for which $\ln \mathcal{L}/\mathcal{L}_{max} \geq -0.49$, this interval is represented by the solid part of the curve in Fig.1a. We repeated the analysis for a set of 228 model clusters with redshifts in the range $0.002 \leq z \leq 0.2$, distributed as $z^2 dz$. The scatter over the whole range of redshifts was $\sigma_z = 0.018$, or $0.009$ if we restricted observations to clusters in the range $z \leq 0.12$. There was no systematic trend with redshift.

As examples of how this technique can be applied to real data, we have analysed two Abell clusters, taken more-or-less at random from the literature. The field of the cluster A3040, which has a redshift of $z = 0.093$ (West & Fransden 1981) is shown in Fig.2. The likelihood function and apparent magniutde distribtion for this cluster are shown in Fig.3. Comparing Fig.1 with Fig.3 shows that the likelihood function is somewhat less strongly peaked for real data, which in turn should imply a larger spread of errors in the redshift estimates when the estimator is applied to a number of real clusters.

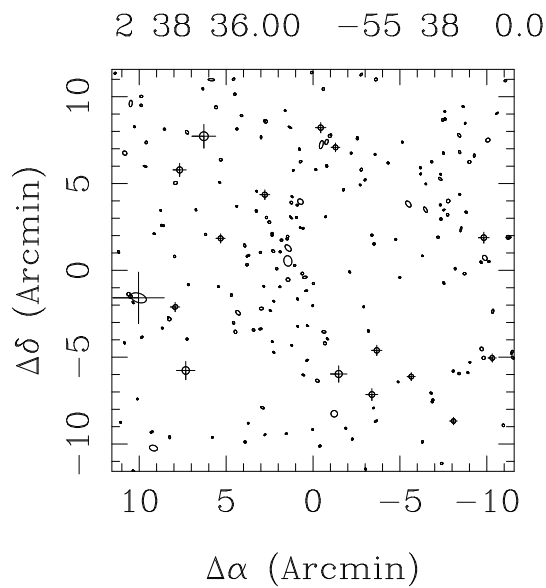

**Figure 2.** The field of view of the cluster A3040 as seen in the APM Survey data. The limiting magnitude of the plot is $b_J = 20.65$. The documented redshift is for the large galaxy in the centre of the field.

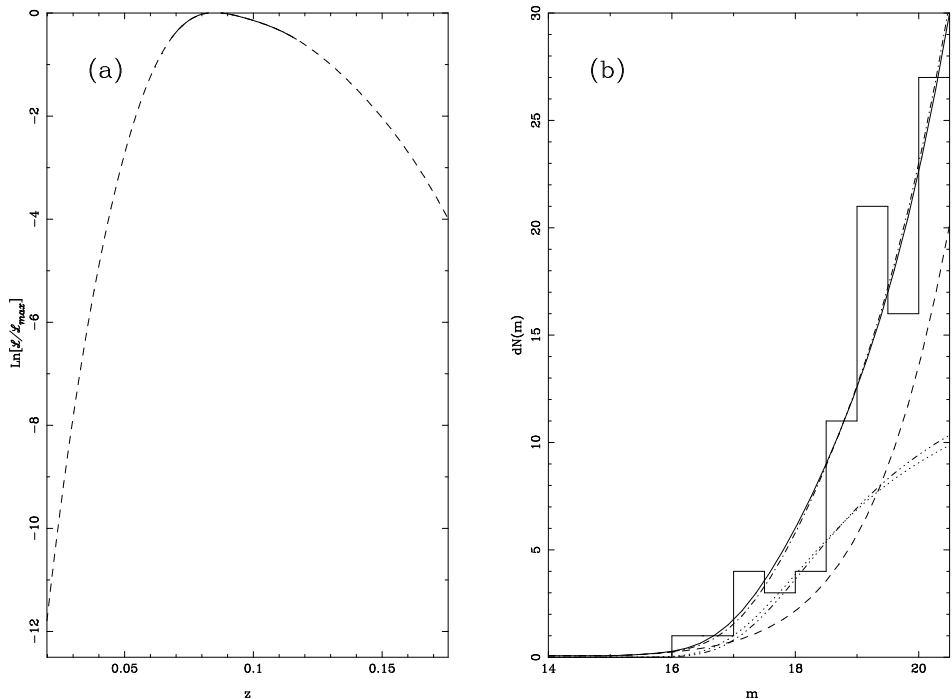

**Figure 3.** a) The likelihood function for A3040. The likelihood function is maximised for $z = 0.0878$. b) The apparent magnitude distribution for A3040. Components corresponding to the model predictions for the field and both the maximum likelihood and listed redshifts are shown as in fig. 1b.

As another example we consider the cluster A2860, listed redshift $z = 0.0268$ (Abell, Corwin & Olowin 1989). The likelihood function (Fig.4a) for this cluster peaks beyond $z = 0.1$, and agrees well with our multi-fibre observations as described in the next section. Fig.5 shows the field of this cluster. Of the 12 galaxies for which good spectra were obtained, 10 have redshifts $z \sim 0.1$. The model predictions for a cluster at $z = 0.0268$ are shown in Fig.4b. The likelihood ratio test gives zero probability for the field to represent a cluster at this redshift. Using our observed redshift as input gives a maximum likelihood redshift estimate of $z_\mathcal{L} = 0.0975$, for which our observed $z = 0.106$ gives a likelihood ratio of 0.94 corresponding to a 71% probability for the field to contain a cluster this redshift.

### 3.4 Application

We apply the maximum likelihood estimator to each cluster interactively. The field of each cluster is initially set to correspond to $0.75\ h^{-1}$Mpc at the redshift to be tested. The value of the background normalisation is defined by the ratio of the observed counts in annuli out to $1.5°$ to the galaxy number counts inferred from our model for the field component. If the redshift of maximum likelihood suggests that the galaxy in question is foreground to the cluster then we repeat the analysis using the maximum likelihood redshift to obtain a new projected radius for the field.

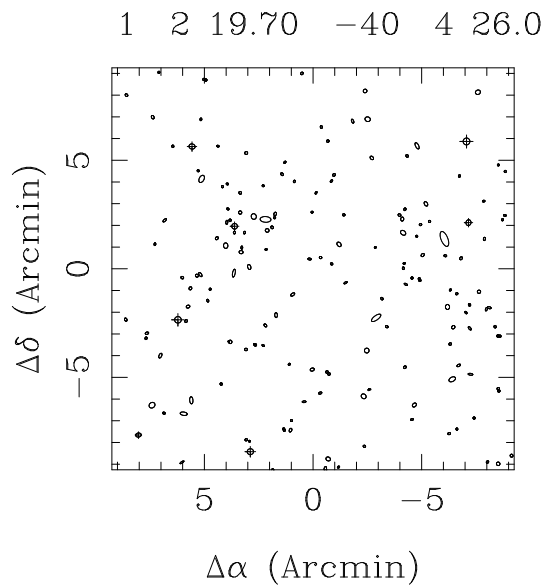

**Figure 5.** The field of view of the cluster A2860 as seen in the APM Survey data. The limiting magnitude of the plot is $b_J = 20.6$. 10 of the brighter galaxies in this field have redshifts $z \sim 0.106$.

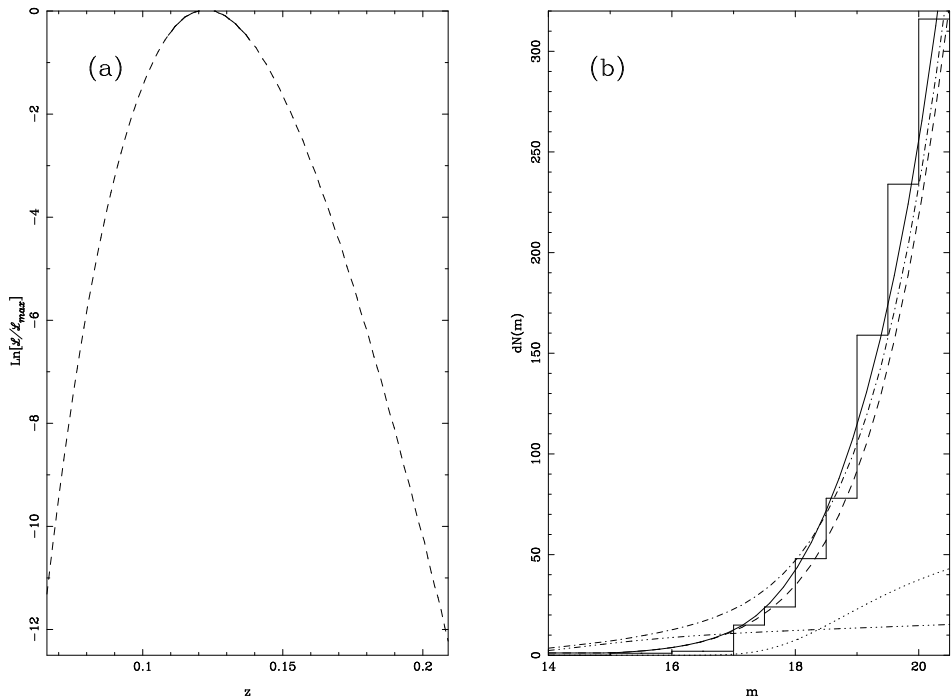

**Figure 4.** a) The likelihood function for A2860 generated using the galaxy distribution within a circle of projected radius $r_C = 0.75\ h^{-1}$ Mpc at the listed redshift of $z = 0.0268$. b) The apparent magnitude distribution for A2860 showing the model predictions for both the maximum likelihood and input redshifts.

In some cases the likelihood function is found to have more than one peak due to the presence of a single very bright bright galaxy. We therefore test each cluster for contamination of the likelihood function by bright galaxies by successively removing the brightest galaxies from the field to obtain a stable redshift estimate.

## 4 EXISTING REDSHIFT DATA

From our cluster catalogue, we selected a target sample of 240 clusters with $\mathcal{R} \geq 20$ and photometrically derived redshifts $z_X \leq 0.1$. Our aim was to determine redshifts for all of these clusters. We began by cross-correlating all 122 clusters from Andernach's compilation which were found within the APM Survey region. We applied the maximum likelihood analysis of §3 to each of these clusters to give an estimate of the cluster redshift in each case, and the likelihood ratio test was used to reject documented redshifts as described above. Where positional information was available from the literature we also inspected the galaxies for which redshifts had been obtained, and rejected all those which appeared to be foreground objects such as bright spirals or galaxies far from the cluster centres.

We cross-referenced the resulting 'decontaminated' list with our target catalogue and obtained 36 redshifts. These are listed in Table 1. We stress that these are not the only Abell clusters with reliable documented redshifts in the APM Survey region, but simply those which are matched to clusters in our target sample. Five clusters in this table have no identification in our catalogue, but are retained as they represent nearby systems which are at low contrast on the sky, and so are missed by the initial stage of the cluster selection process. The relationship between redshift and $m_X$ for the matched clusters is shown in Fig. 6. The best fit relation is

$$\log z = -3.523 + 0.133 m_X$$

with a scatter of 0.08 in $\log z$.

## 5 THE REDSHIFT SURVEY

Our initial observing strategy was to obtain redshifts using a multiplexed setup to observe at least 10 galaxies per cluster. We observed 23 APM clusters in Octber 1989 using the AUTOFIB system and Image Photon Counting System (IPCS) at the Anglo-Australian Telescope (AAT) with wavelength coverage of $3670\text{Å}$-$5680\text{Å}$ and $2\text{Å}$/pixel, giving redshifts for $\gtrsim 20$ galaxies per cluster in 3 nights of observing. Redshifts were determined from a sample of stellar templates using the cross-correlation technique of Tonry & Davis (1979), with a lower threshold for acceptance of the redshift set at $r = 2.5$, where $r$ is the signal to noise ratio of the peak in the cross-correlation function as defined by Tonry & Davis. In a few cases galaxy redshifts were determined from emission line features, but these are rare given that we restricted our observations to early-type galaxies. The AUTOFIB clusters are listed in table ??. Entries in this table which have no entries for $\mathcal{R}$ or $m_X$ were included in the original selection for this run, and were retained in our final sample for the same reasons as the nearby Abell clusters from table 1.

**Table 1.** Clusters for which redshifts were found in Andernach's compilation after rejection of probable redshift misidentifications. Those clusters with no entries in columns 3 and 4 are nearby systems which could be missed in the percolation stage of our cluster selection procedure.

| (1) $\alpha$(1950) | (2) $\delta$(1950) | (3) $m_X$ | (4) $\mathcal{R}$ | (5) $z$ | (6) Abell |
|---|---|---|---|---|---|
| 00 03 43.51 | -34 59 05.27 | 19.083 | 69.943 | 0.116 | A2721 |
| 00 18 01.94 | -49 33 47.53 | 17.608 | 30.170 | 0.064 | A2764 |
| 00 18 04.13 | -25 58 36.48 | 19.396 | 72.301 | 0.131 | A0022 |
| 00 23 00.67 | -33 19 18.12 | 16.701 | 30.030 | 0.050 | S0041 |
| 00 26 07.54 | -23 53 19.32 | 19.227 | 45.402 | 0.109 | A0042 |
| 01 00 19.56 | -22 09 07.92 | 17.988 | 38.168 | 0.060 | A0133 |
| 01 07 40.08 | -46 10 28.56 | 15.983 | 31.139 | 0.023 | A2877 |
| 02 49 15.29 | -25 07 54.84 | 18.654 | 74.823 | 0.116 | A0389 |
| 03 06 03.12 | -23 52 49.44 | — | — | 0.041 | A0419 |
| 03 14 51.84 | -51 05 56.40 | 17.940 | 31.065 | 0.075 | A3110 |
| 03 16 09.82 | -44 25 16.68 | 18.258 | 49.031 | 0.072 | A3112 |
| 03 17 54.00 | -54 02 60.00 | — | — | 0.055 | S0339 |
| 03 25 59.02 | -53 53 06.72 | 17.017 | 39.791 | 0.059 | A3125 |
| 03 27 23.50 | -55 52 41.88 | 18.766 | 91.874 | 0.086 | A3126 |
| 03 29 07.94 | -52 43 04.08 | 17.264 | 59.761 | 0.059 | A3128 |
| 03 39 05.30 | -55 13 12.00 | — | — | 0.043 | S0377 |
| 03 41 42.29 | -53 47 57.84 | 18.082 | 62.224 | 0.058 | A3158 |
| 03 43 38.88 | -24 25 58.44 | 19.110 | 48.669 | 0.105 | A0458 |
| 04 30 31.78 | -61 31 51.96 | 18.120 | 47.493 | 0.059 | A3266 |
| 04 36 36.29 | -22 14 26.16 | 17.560 | 47.456 | 0.067 | A0500 |
| 04 46 01.44 | -20 33 14.40 | 17.819 | 45.751 | 0.073 | A0514 |
| 04 59 03.99 | -22 53 01.32 | — | — | 0.047 | A0533 |
| 20 35 36.36 | -61 24 36.36 | 16.742 | 43.854 | 0.071 | A3703 |
| 20 38 44.14 | -35 25 40.81 | 18.300 | 62.379 | 0.090 | A3705 |
| 20 48 07.92 | -52 56 04.92 | 15.741 | 44.932 | 0.047 | S0906 |
| 21 31 03.56 | -53 51 02.52 | 17.591 | 32.325 | 0.078 | A3785 |
| 21 49 31.83 | -19 48 40.32 | 19.327 | 69.692 | 0.094 | A2384 |
| 21 58 17.90 | -60 11 05.63 | 18.470 | 40.567 | 0.099 | A3827 |
| 22 17 02.14 | -55 28 18.84 | — | — | 0.040 | A3869 |
| 22 21 29.66 | -64 30 37.44 | 18.967 | 39.439 | 0.094 | S1022 |
| 22 30 05.18 | -55 03 49.33 | 18.779 | 41.799 | 0.075 | A3886 |
| 22 59 33.70 | -22 17 02.76 | 19.338 | 50.752 | 0.136 | A2521 |
| 23 02 54.94 | -21 38 42.36 | 18.295 | 35.898 | 0.095 | A2528 |
| 23 05 55.22 | -20 09 28.44 | 18.780 | 51.898 | 0.083 | A2538 |
| 23 09 36.52 | -21 50 16.80 | 18.809 | 61.707 | 0.086 | A2556 |
| 23 56 20.59 | -60 55 55.20 | 19.318 | 47.358 | 0.096 | A4067 |

The data from these observations, together with additional multi-object data from Colless & Hewett (1987) and Teague, Carter & Gray (1990) suggested that restricting observations to the brightest pair of early-type galaxies within our cluster defining radius would give the correct cluster redshifts with only a small number of contaminants. The data are shown in figure 7. Applying the likelihood ratio test of §3 to these clusters rejected only those redshifts which are shown as open symbols. This shows that the likelihood ratio test succeeds in rejecting all foreground galaxies. The residual scatter of the solid points in figure 7 is 512km s$^{-1}$.

We therefore adopted as optimal the strategy of observing only one pair of galaxies for each remaining cluster using a single slit oriented to observe both galaxies simultaneously.

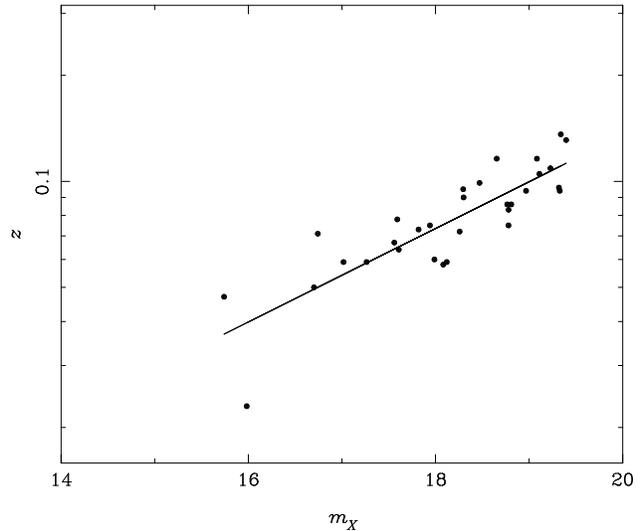

**Figure 6.** The relationship between redshift and $m_X$ for the sample of clusters with documented redshifts.

**Table 2.** Clusters observed using AUTOFIB. Column (6) gives the number of galaxies for which redshifts were obtained in each field. Clusters with no entries in columns (3) or (4) were included in the original selection for the AUTOFIB observations and are included in our final list as nearby systems (see text).

| (1) $\alpha$(1950) | (2) $\delta$(1950) | (3) $m_X$ | (4) $\mathcal{R}$ | (5) $z$ | (6) $N_z$ | (7) Abell |
|---|---|---|---|---|---|---|
| 00 35 04.8 | -28 47 53 | 18.918 | 34.944 | 0.113 | 21 | A2798 |
| 00 46 53.9 | -29 47 54 | 19.011 | 51.255 | 0.108 | 18 | S0084 |
| 01 02 19.7 | -40 04 26 | 18.096 | 34.959 | 0.106 | 12 | A2860 |
| 01 12 00.9 | -32 06 26 | 15.714 | 21.800 | 0.020 | 14 | S0141 |
| 01 29 24.6 | -51 37 9 | 16.260 | 22.095 | 0.055 | 19 | S0162 |
| 01 39 44.7 | -42 23 32 | 17.587 | 30.208 | 0.076 | 15 | S0180 |
| 02 28 32.8 | -33 19 32 | 18.389 | 37.181 | 0.076 | 21 | A3027 |
| 03 09 42.7 | -53 13 26 | — | — | 0.053 | 18 |  |
| 03 35 37.1 | -55 10 46 | 16.697 | 25.683 | 0.045 | 16 | A3144 |
| 04 52 31.0 | -18 22 36 | — | — | 0.030 | 16 |  |
| 20 57 24.6 | -38 47 54 | — | — | 0.046 | 17 | A3733 |
| 20 58 43.6 | -28 18 23 | 16.533 | 26.885 | 0.038 | 22 |  |
| 20 59 36.7 | -41 34 46 | 16.527 | 26.950 | 0.082 | 16 | A3739 |
| 21 08 08.7 | -23 20 32 | 15.867 | 29.690 | 0.033 | 15 |  |
| 21 46 34.5 | -55 35 29 | 16.482 | 25.542 | 0.036 | 17 | A3816 |
| 21 50 18.2 | -55 50 49 | — | — | 0.036 | 17 | A3816 |
| 21 59 39.3 | -22 50 22 | 17.501 | 53.990 | 0.070 | 15 | S0987 |
| 22 15 39.0 | -39 08 17 | 19.315 | 44.986 | 0.126 | 2 | A3856 |
| 22 24 19.2 | -30 49 1 | 17.140 | 37.057 | 0.057 | 10 | A3880 |
| 22 34 18.7 | -38 16 49 | 19.337 | 37.279 | 0.105 | 10 | S1045 |
| 23 28 01.3 | -35 20 21 | 16.638 | 21.957 | 0.050 | 17 | A4013 |
| 23 44 19.7 | -28 27 16 | 16.173 | 22.113 | 0.029 | 30 | A4038 |
| 23 53 35.5 | -34 39 55 | — | — | 0.048 | 11 | A4059 |

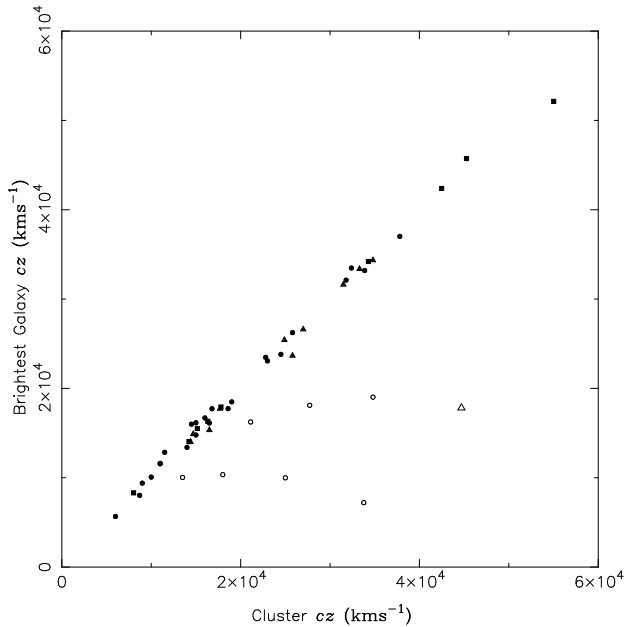

**Figure 7.** The redshifts of the brightest galaxies in cluster fields relative to the true cluster redshifts. The data shown are from Colless & Hewett (1987) (triangles), Teague, Carter & Gray (1990) (squares) and this work (circles). Open symbols show those points which were rejected by the likelihood ratio test as described in § 3.

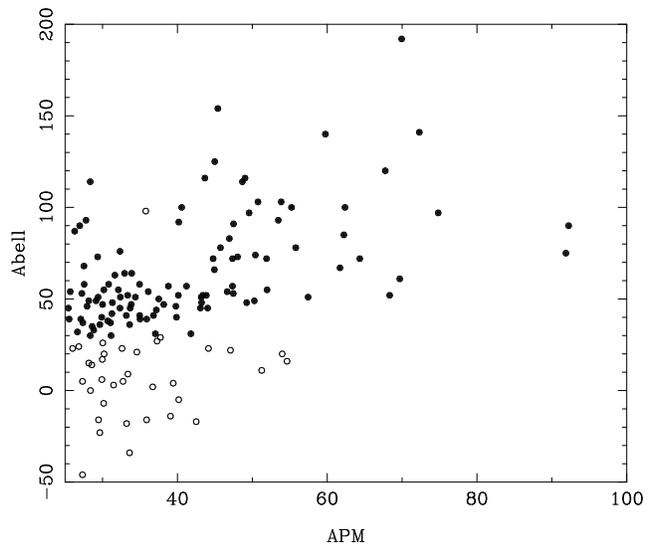

**Figure 10.** A comparison of our richness estimates, $\mathcal{R}$, with the Abell counts for clusters with Abell identifications in Table 4. Open circles denote supplementary Abell clusters.

In many cases this procedure allowed observations for more than two galaxies per cluster subject to the constraint of possible slit alignments.

We obtained redshifts for 141 clusters at the AAT in four nights in November of 1990, using the RGO spectrograph, $600V$ grating, and IPCS, giving 2048 bins in the wavelength direction with a resolution of $\sim 1\text{Å}/\text{pixel}$ in the range $3670$–$5680\text{Å}$ and a useable slit length of $3.5'$. Sample spectra, showing the absorbtion and emission features used to determine the redshifts are shown in figure 8. redshifts for a further 32 clusters were obtiained by Jon Loveday using the Mt. Stromlo 2.3m telescope. Before observing we cross-referenced our list of target galaxies with Huchra (1990), which gave another 5 cluster redshifts. A further four redshifts were rejected by the maximum likelihood analysis, showing that carefully selecting the galaxies to be observed gives a significant improvement over simply selecting the brightest galaxy in the cluster field independently of position relative to the cluster centre.

In table 3 we list the galaxy redshifts for each cluster, together with the maximum likelihood redshift estimate, and the probability for the observed galaxy to be a cluster member. The numbers in columns 3 and 4 refer to the whole magnitude distribution of the cluster field. We also checked each likelihood interactively to remove any possible contamination from bright galaxies. Where changes were necessary the adopted values of the maximum likelihood redshift and probability for cluster membership are given in columns 5 and 6.

In §3 we assumed that the luminosity function for galaxies in clusters was the same as that for the field. We tested this assumption by changing the input luminosity function parameters to

$$B_{J*} = -20.05; \quad \alpha = -1.25,$$

consistent with the findings of Rhee (1989), Colless (1989), and with our own internal estimates based on this sample of clusters (*in preparation*). We find that the changes in the likelihood estimates obtained in this way are within the original error estimates, and that repeating the interactive analysis with these parameters produces no change to our cluster redshift list.

The cluster redshift list is given in Table 4. The distribution of our survey in redshift-space is shown in figure 9. Where the centre of an Abell cluster lies within the counting circle of an APM cluster we give the Abell identification, although in some cases our use of a small defining radius means that an Abell cluster will be identified as more than one APM cluster. Fig. 10 shows a large scatter between our richness measure, $\mathcal{R}$ and the Abell richness listed by Abell, Corwin & Olowin's (1989). We note that a considerable number of our clusters identify with the supplementary clusters of ACO. This is not surprising given the large scatter in the Abell richness counts (see also Lumsden *et al.* 1992). A more detailed comparison between our catalog and the ACO catalog, with a discussion of possible error mechanisms within the Abell catalogues will be given in a future paper.

## 6 SUMMARY

We have selected a catalogue of rich galaxy clusters from the APM Galaxy Survey using uniform selection criteria. Using redshifts from the literature and a small number of multi-fibre observations we have optimised our observing strategy and obtained a large, complete sample of cluster redshifts by making only one single-slit observation of most of our clusters. Foreground redshifts have been removed from the final sample by using a maximum likelihood analysis of the APM data for each cluster field. The two-point correlation function for this sample has been discussed by Dalton *et al.*

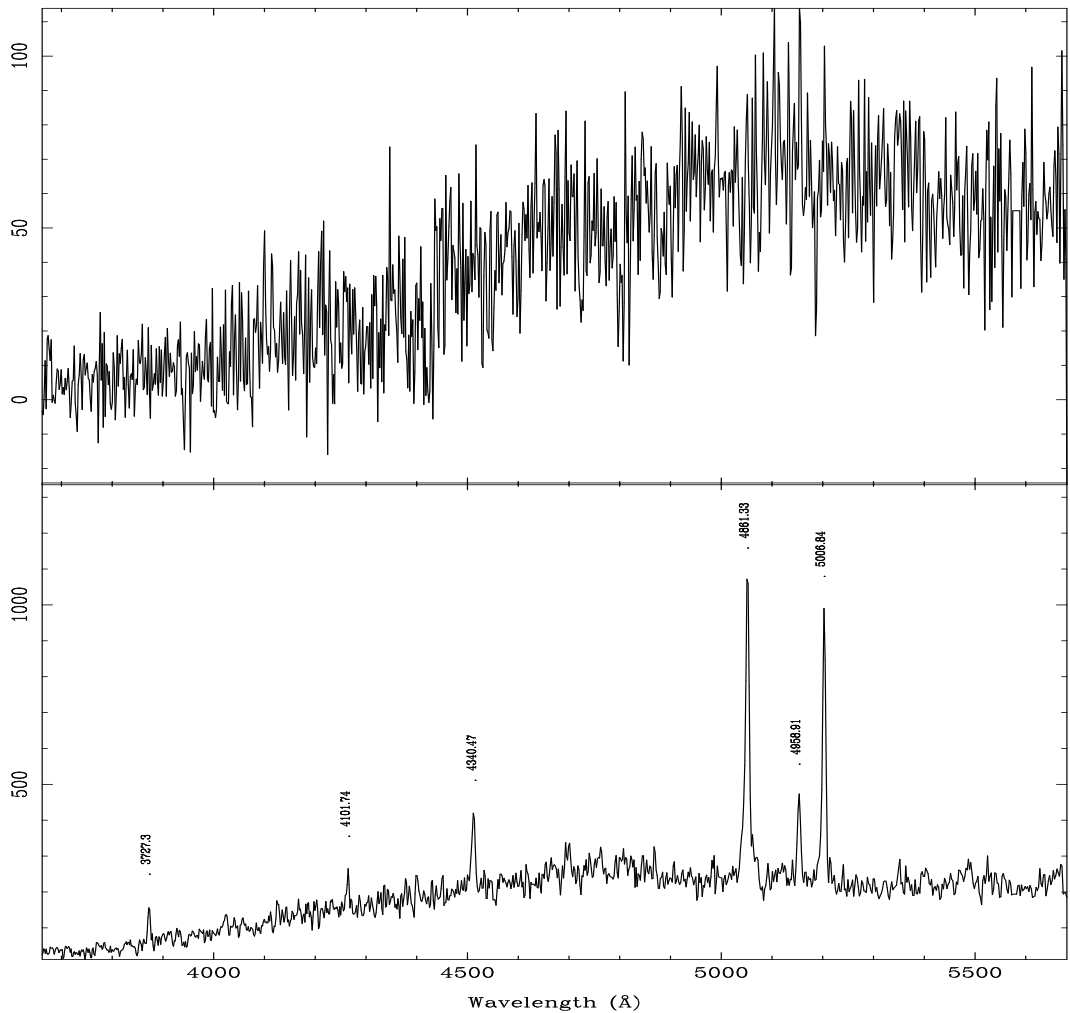

**Figure 8.** Typical spectra showing the absorbtion, and emission features used for redshift determinations.

(1992), and a comparison of the clustering properties of this sample with redshift samples of Abell clusters has been presented by Efstathiou *et al.* (1992).

An extension of the redshift survey is currently in progress, with the intention of obtaining redshifts for over 500 rich clusters using similar techniques to those described here. We will discuss the extension, together with modifcations of the cluster finding algorithm and detailed comparisons with the Abell catalogue in future papers.

## 7  ACKNOWLEDGEMENTS

We would like to thank the staff of the Anglo-Australian Telescope, particularly Ray Sharples for technical assistance with AUTOFIB, and Max Pettini for observing support. This work has been supported by grants from the UK Science and Engineering Research Council. G.B.D acknowledges the receipt of an SERC studentship and travel grants in support of this work.

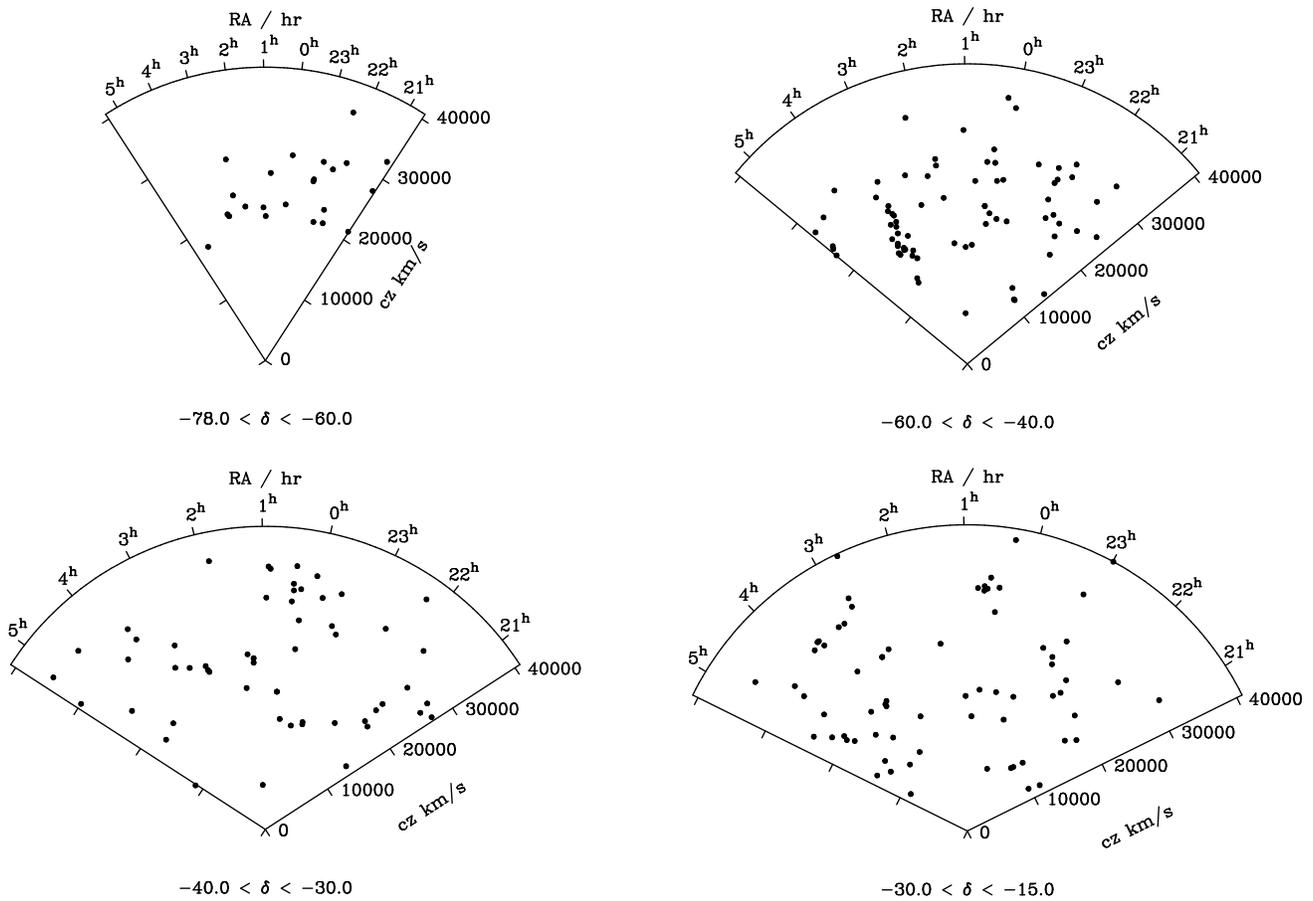

**Figure 9.** The sky distribution of the survey. The large structure seen at $\alpha = 3$ hours is the Horologium-Reticulum supercluster.

**Table 3.** Redshifts for galaxies in clusters observed using a single slit. Entries in columns (5) and (6) only occur for those fields where an interactive application of the analysis changed the redshift of maximum likelihood (see text).

| (1) Cluster | (2) $z_{obs}$ | (3) $z_{\mathcal{L}}$ | (4) $P$ | (5) $z_{\mathcal{L}int}$ | (6) $P_{int}$ |
|---|---|---|---|---|---|
| 0001-5103 | 0.0393 | 0.1419 | 0.002* | | |
| 0001-5103 | 0.1179 | 0.1213 | 0.81 | 0.1179 | 1.00 |
| 0008-2908 | 0.0602 | 0.0555 | 0.5708 | | |
| 0008-2908 | 0.0646 | 0.0543 | 0.3090* | | |
| 0009-4232 | 0.0860 | 0.0908 | 0.84 | 0.0911 | 0.42 |
| 0009-4232 | 0.0860 | 0.0956 | 0.50 | 0.0907 | 0.69 |
| 0011-4317 | 0.1199 | 0.0821 | 0.0998* | 0.1027 | 0.4114 |
| 0011-4317 | 0.1234 | 0.0773 | 0.0664* | 0.0986 | 0.2690* |
| 0013-3136 | 0.0719 | 0.0981 | 0.6439 | | |
| 0013-3136 | 0.0805 | 0.1020 | 0.5557 | | |
| 0013-4850 | 0.0677 | 0.0497 | 0.41 | 0.0821 | 0.54 |
| 0013-4850 | 0.0699 | 0.0531 | 0.53 | 0.0923 | 0.40 |
| 0013-4850 | 0.0700 | 0.0531 | 0.53 | 0.0925 | 0.38 |
| 0015-3526 | 0.0903 | 0.0803 | 0.30* | 0.0877 | 0.73 |
| 0015-3526 | 0.0946 | 0.0801 | 0.15* | 0.0895 | 0.41 |
| 0015-3526 | 0.0948 | 0.0802 | 0.15* | 0.0895 | 0.41 |
| 0018-3413 | 0.1069 | 0.1039 | 0.83 | | |
| 0018-3413 | 0.1087 | 0.1025 | 0.69 | | |
| 0018-3413 | 0.1088 | 0.1026 | 0.69 | | |
| 0018-3413 | 0.1103 | 0.1009 | 0.61 | | |
| 0020-5353 | 0.0966 | 0.1020 | 0.73 | | |
| 0020-5353 | 0.0985 | 0.1040 | 0.69 | | |
| 0024-4849 | 0.0720 | 0.1030 | 0.4206 | | |
| 0024-4907 | 0.0710 | 0.0939 | 0.25* | 0.0939 | 0.39 |
| 0024-4907 | 0.0722 | 0.0955 | 0.27* | 0.0955 | 0.40 |
| 0025-3033 | 0.1135 | 0.1200 | 0.83 | 0.1200 | 0.68 |
| 0025-3033 | 0.1192 | 0.1192 | 1.00 | 0.1209 | 0.90 |
| 0025-3542 | 0.1067 | 0.1097 | 0.86 | | |
| 0025-3542 | 0.1076 | 0.1076 | 1.00 | | |
| 0025-3542 | 0.1078 | 0.1078 | 1.00 | | |
| 0026-3032 | 0.0744 | 0.0985 | 0.16* | | |
| 0026-3032 | 0.1032 | 0.1061 | 0.87 | | |
| 0026-3516 | 0.1106 | 0.1043 | 0.50 | 0.1156 | 0.69 |
| 0026-3516 | 0.1109 | 0.1014 | 0.46 | 0.1156 | 0.69 |
| 0027-2944 | 0.0657 | 0.0831 | 0.3689 | | |
| 0027-2944 | 0.0972 | 0.0808 | 0.3546 | | |
| 0027-2944 | 0.0979 | 0.0787 | 0.3293 | | |
| 0027-5341 | 0.0862 | 0.1100 | 0.04* | | |
| 0027-5341 | 0.0864 | 0.1103 | 0.04* | | |
| 0027-5341 | 0.0921 | 0.1075 | 0.16* | 0.1023 | 0.34 |
| 0035-3107 | 0.0606 | 0.1365 | 0.000* | | † |
| 0035-3107 | 0.0622 | 0.1340 | 0.001* | | † |
| 0035-3107 | 0.0626 | 0.1349 | 0.001* | | † |
| 0035-3924 | 0.0622 | 0.0622 | 1.00 | | |
| 0035-3924 | 0.0623 | 0.0623 | 1.00 | | |
| 0035-3924 | 0.0633 | 0.0666 | 0.66 | | |
| 0036-2234 | 0.0631 | 0.0565 | 0.6335 | | |
| 0036-2234 | 0.0645 | 0.0577 | 0.6027 | | |
| 0037-2625 | 0.0759 | 0.1089 | 0.3007* | 0.1007 | 0.4425 |
| 0040-2621 | 0.1067 | 0.0901 | 0.24* | 0.1052 | 0.88 |
| 0040-2621 | 0.1089 | 0.0904 | 0.19* | 0.1058 | 0.78 |
| 0040-2621 | 0.1115 | 0.0956 | 0.25* | 0.1083 | 0.78 |
| 0040-2852 | 0.1069 | 0.0902 | 0.23* | 0.0978 | 0.43 |
| 0040-2852 | 0.1094 | 0.0877 | 0.14* | | |
| 0043-6351 | 0.0746 | 0.0807 | 0.35 | 0.0826 | 0.24* |
| 0043-6351 | 0.0750 | 0.0791 | 0.43 | 0.0811 | 0.29* |
| 0043-6351 | 0.0866 | 0.0794 | 0.28* | 0.0806 | 0.34 |

**Table 3.** (*continued*)

| (1) Cluster | (2) $z_{obs}$ | (3) $z_{\mathcal{L}}$ | (4) $P$ | (5) $z_{\mathcal{L}int}$ | (6) $P_{int}$ |
|---|---|---|---|---|---|
| 0044-5500 | 0.0811 | 0.0855 | 0.67 | | |
| 0044-5500 | 0.0830 | 0.0853 | 0.76 | | |
| 0045-6334 | 0.0305 | 0.1074 | 0.00* | 0.1123 | 0.00* |
| 0045-6334 | 0.0593 | 0.0963 | 0.01* | 0.1086 | 0.001* |
| 0046-4215 | 0.0533 | 0.1507 | 0.02* | 0.0587 | 0.63 |
| 0046-4215 | 0.0537 | 0.2667 | 0.01* | 0.0592 | 0.60 |
| 0048-2846 | 0.0510 | 0.1125 | 0.00* | 0.0664 | 0.33 |
| 0048-2846 | 0.0550 | 0.1113 | 0.00* | 0.0662 | 0.50 |
| 0051-3117 | 0.0287 | 0.0287 | 1.00 | 0.1070 | 0.00* |
| 0051-3117 | 0.1170 | 0.1070 | 0.62 | | |
| 0054-3809 | 0.1154 | 0.1022 | 0.52 | | |
| 0054-3809 | 0.1175 | 0.1007 | 0.46 | | |
| 0056-3432 | 0.0878 | 0.1267 | 0.0710* | | |
| 0056-3432 | 0.1040 | 0.1267 | 0.3619 | | |
| 0056-6704 | 0.0663 | 0.0857 | 0.23* | 0.0734 | 0.52 |
| 0056-6704 | 0.0683 | 0.0829 | 0.19* | 0.0756 | 0.35 |
| 0101-4307 | 0.0506 | 0.0519 | 0.84 | | |
| 0101-4307 | 0.0548 | 0.0548 | 1.00 | | |
| 0102-6710 | 0.0697 | 0.0660 | 0.80 | 0.0697 | 1.00 |
| 0102-6710 | 0.0706 | 0.0668 | 0.78 | 0.0687 | 0.85 |
| 0115-3650 | 0.0734 | 0.1130 | 0.20* | | |
| 0115-3650 | 0.0752 | 0.1119 | 0.32* | | |
| 0115-3815 | 0.0772 | 0.0772 | 1.00 | | |
| 0115-3815 | 0.0773 | 0.0773 | 1.00 | | |
| 0115-3815 | 0.1189 | 0.0730 | 0.03* | | |
| 0124-3810 | 0.0773 | 0.1046 | 0.02* | 0.0941 | 0.33 |
| 0124-3810 | 0.0807 | 0.1028 | 0.06* | 0.0939 | 0.46 |
| 0131-2714 | 0.0821 | 0.1182 | 0.17* | | |
| 0131-2714 | 0.0838 | 0.1207 | 0.20* | | |
| 0132-2740 | 0.0869 | 0.1782 | 0.00* | | |
| 0132-2740 | 0.0873 | 0.1791 | 0.00* | | |
| 0132-3305 | 0.0638 | 0.0571 | 0.73 | 0.0722 | 0.47 |
| 0132-3305 | 0.0646 | 0.0492 | 0.58 | 0.0714 | 0.56 |
| 0144-5539 | 0.0944 | 0.0918 | 0.82 | | |
| 0144-5539 | 0.1346 | 0.0879 | 0.004* | | |
| 0144-5618 | 0.0906 | 0.1284 | 0.08* | | |
| 0144-5618 | 0.0920 | 0.1330 | 0.09* | | |
| 0152-3555 | 0.0334 | 0.1270 | 0.001* | | |
| 0152-3555 | 0.1228 | 0.1158 | 0.67 | | |
| 0156-6438 | 0.0696 | 0.0938 | 0.11* | 0.0845 | 0.35 |
| 0156-6438 | 0.0735 | 0.0933 | 0.18* | 0.0893 | 0.42 |
| 0159-4829 | 0.0853 | 0.0924 | 0.56 | | |
| 0159-4829 | 0.0873 | 0.0873 | 1.00 | | |
| 0221-4840 | 0.0738 | 0.0698 | 0.84 | | |
| 0221-4840 | 0.0747 | 0.0707 | 0.81 | | |
| 0225-6714 | 0.0942 | 0.1179 | 0.17* | 0.1152 | 0.51 |
| 0225-6714 | 0.0968 | 0.1185 | 0.20* | 0.1076 | 0.59 |
| 0225-6956 | 0.0772 | 0.0793 | 0.90 | 0.0856 | 0.65 |
| 0225-6956 | 0.0781 | 0.0781 | 1.00 | 0.0868 | 0.67 |
| 0227-3404 | 0.0730 | 0.0966 | 0.24* | | |
| 0227-3404 | 0.0751 | 0.0975 | 0.24* | | |
| 0227-3404 | 0.0810 | 0.0987 | 0.35 | | |
| 0229-2311 | 0.0550 | 0.0550 | 1.00 | 0.055 | 1.00 |
| 0229-2311 | 0.0557 | 0.0557 | 1.00 | 0.0557 | 1.00 |
| 0229-2311 | 0.0573 | 0.0573 | 1.00 | 0.0573 | 1.00 |
| 0229-3215 | 0.0767 | 0.0934 | 0.20* | 0.0892 | 0.60 |
| 0229-3215 | 0.0780 | 0.0993 | 0.20* | 0.0949 | 0.41 |

Table 3. (*continued*)

| (1) Cluster | (2) $z_{obs}$ | (3) $z_\mathcal{L}$ | (4) $P$ | (5) $z_{\mathcal{L}int}$ | (6) $P_{int}$ |
|---|---|---|---|---|---|
| 0232-5948 | 0.0877 | 0.1023 | 0.31* | | |
| 0232-5948 | 0.0905 | 0.1006 | 0.44 | | |
| 0234-1934 | 0.0861 | 0.0932 | 0.49 | | |
| 0234-1934 | 0.0917 | 0.0917 | 1.00 | 0.1019 | 0.41 |
| 0242-2625 | 0.1350 | 0.0940 | 0.01* | 0.1271 | 0.49 |
| 0242-2625 | 0.1464 | 0.0910 | 0.000* | | |
| 0245-1958 | 0.0858 | 0.0882 | 0.86 | | |
| 0245-1958 | 0.0873 | 0.0873 | 1.00 | | |
| 0245-2250 | 0.0838 | 0.0638 | 0.5066 | | |
| 0245-2250 | 0.0857 | 0.0644 | 0.4514 | | |
| 0245-2250 | 0.0859 | 0.0645 | 0.4491 | | |
| 0245-2250 | 0.0884 | 0.0688 | 0.3867 | | |
| 0249-2549 | 0.1117 | 0.1117 | 1.00 | | |
| 0249-2549 | 0.1198 | 0.1061 | 0.59 | | |
| 0249-7136 | 0.0676 | 0.1252 | 0.0564* | | † |
| 0249-7136 | 0.0689 | 0.1278 | 0.0622* | | † |
| 0249-7136 | 0.0705 | 0.1309 | 0.0413* | | † |
| 0253-3537 | 0.0794 | 0.0837 | 0.76 | 0.0924 | 0.53 |
| 0253-3537 | 0.0799 | 0.0843 | 0.78 | 0.0930 | 0.53 |
| 0253-3537 | 0.0815 | 0.0860 | 0.79 | 0.0949 | 0.53 |
| 0253-6636 | 0.0702 | 0.0740 | 0.75 | 0.0777 | 0.56 |
| 0253-6636 | 0.0714 | 0.0733 | 0.84 | 0.0791 | 0.60 |
| 0258-3638 | 0.0494 | 0.0691 | 0.17* | | |
| 0258-3638 | 0.0917 | 0.0777 | 0.49 | 0.0935 | 0.70 |
| 0258-3638 | 0.0958 | 0.0770 | 0.36 | 0.0958 | 1.00 |
| 0304-1752 | 0.1062 | 0.1212 | 0.40 | 0.1182 | 0.12* |
| 0304-1752 | 0.1062 | 0.1212 | 0.40 | 0.1182 | 0.12* |
| 0304-1752 | 0.1069 | 0.1190 | 0.43 | 0.1159 | 0.64 |
| 0307-4727 | 0.0626 | 0.0922 | 0.0586* | 0.0791 | 0.3866 |
| 0307-4727 | 0.0638 | 0.0907 | 0.0912* | 0.0726 | 0.4832 |
| 0309-2707 | 0.0684 | 0.0684 | 1.00 | | |
| 0309-2707 | 0.0684 | 0.0684 | 1.00 | | |
| 0309-2707 | 0.1664 | 0.0491 | 0.00* | | |
| 0310-2721 | 0.0643 | 0.0575 | 0.5630 | | |
| 0310-2721 | 0.0646 | 0.0646 | 1.00 | | |
| 0310-2721 | 0.1074 | 0.0557 | 0.0398* | | |
| 0310-2721 | 0.1080 | 0.0682 | 0.07* | | |
| 0310-5305 | 0.0570 | 0.1862 | 0.0005* | | † |
| 0311-3829 | 0.0798 | 0.1016 | 0.15* | | |
| 0311-3829 | 0.0847 | 0.1010 | 0.28* | 0.1034 | 0.39 |
| 0313-1917 | 0.0657 | 0.0936 | 0.15* | 0.0831 | 0.34 |
| 0313-1917 | 0.0676 | 0.0892 | 0.20* | 0.0819 | 0.40 |
| 0313-2926 | 0.0641 | 0.0607 | 0.7553 | | |
| 0313-2926 | 0.0670 | 0.0598 | 0.5888 | | |
| 0314-4406 | 0.0916 | 0.0865 | 0.7203 | | |
| 0315-4442 | 0.0760 | 0.0884 | 0.27* | 0.0842 | 0.38 |
| 0315-4442 | 0.1287 | 0.1064 | 0.26* | | |
| 0316-4551 | 0.0766 | 0.0766 | 1.00 | | |
| 0316-4551 | 0.0767 | 0.0767 | 1.00 | | |
| 0316-4551 | 0.0808 | 0.0808 | 1.00 | | |
| 0316-4551 | 0.0821 | 0.0776 | 0.71 | | |
| 0320-2459 | 0.0857 | 0.1330 | 0.07* | | † |
| 0320-2459 | 0.0871 | 0.1305 | 0.07* | | † |
| 0320-2459 | 0.0884 | 0.1276 | 0.09* | | † |
| 0320-4544 | 0.0670 | 0.0706 | 0.78 | | |
| 0320-4544 | 0.0696 | 0.0733 | 0.76 | | |
| 0320-4544 | 0.0717 | 0.0717 | 1.00 | | |
| 0320-5320 | 0.0768 | 0.0664 | 0.50 | 0.0872 | 0.36 |
| 0320-5320 | 0.0769 | 0.0664 | 0.49 | 0.0874 | 0.37 |
| 0320-5320 | 0.0798 | 0.0645 | 0.33 | 0.0863 | 0.60 |
| 0320-5320 | 0.0798 | 0.0645 | 0.33 | 0.0863 | 0.60 |
| 0320-5320 | 0.0799 | 0.0646 | 0.33 | 0.0843 | 0.63 |
| 0323-5845 | 0.0655 | 0.0620 | 0.73 | | |
| 0323-5845 | 0.0684 | 0.0611 | 0.57 | | |
| 0327-4610 | 0.0702 | 0.0627 | 0.49 | | |
| 0327-4610 | 0.0720 | 0.0623 | 0.36 | | |
| 0333-2900 | 0.0423 | 0.0423 | 1.00 | 0.1071 | 0.00* |
| 0333-2900 | 0.1037 | 0.1037 | 1.00 | | |
| 0334-2812 | 0.1052 | 0.0903 | 0.40 | 0.0874 | 0.39 |
| 0334-2812 | 0.1068 | 0.0947 | 0.48 | 0.0917 | 0.35 |
| 0334-2812 | 0.1486 | 0.0836 | 0.003* | | |
| 0334-5350 | 0.0622 | 0.0622 | 1.00 | | |
| 0334-5350 | 0.0627 | 0.0627 | 1.00 | | |
| 0335-3957 | 0.0674 | 0.0961 | 0.06* | | |
| 0335-3957 | 0.0683 | 0.0975 | 0.07* | | |
| 0335-3957 | 0.1034 | 0.0946 | 0.51 | 0.0990 | 0.66 |
| 0336-2510 | 0.0507 | 0.0507 | 1.00 | 0.0557 | 0.65 |
| 0336-2510 | 0.0533 | 0.0533 | 1.00 | 0.0534 | 1.00 |
| 0336-2510 | 0.0534 | 0.0534 | 1.00 | 0.0534 | 1.00 |
| 0336-2843 | 0.1071 | 0.1071 | 1.00 | | |
| 0336-2843 | 0.1074 | 0.1074 | 1.00 | | |
| 0336-3319 | 0.1088 | 0.1026 | 0.75 | | |
| 0336-3319 | 0.1116 | 0.1116 | 1.00 | | |
| 0336-4045 | 0.0620 | 0.0944 | 0.3750 | | |
| 0338-2850 | 0.0658 | 0.0606 | 0.54 | 0.0658 | 1.00 |
| 0338-2850 | 0.0685 | 0.0612 | 0.39 | 0.0667 | 0.79 |
| 0339-4551 | 0.0661 | 0.0801 | 0.07* | 0.0765 | 0.36 |
| 0339-4551 | 0.0670 | 0.0830 | 0.08* | 0.0741 | 0.42 |
| 0343-4123 | 0.0536 | 0.0754 | 0.11* | 0.0645 | 0.37 |
| 0343-4123 | 0.0589 | 0.0711 | 0.31* | 0.0650 | 0.60 |
| 0345-1748 | 0.0233 | 0.1563 | 0.00* | | |
| 0345-1748 | 0.1486 | 0.1486 | 1.00 | | |
| 0346-1807 | 0.0376 | 0.1618 | 0.00* | 0.1618 | † |
| 0346-1807 | 0.0386 | 0.1601 | 0.00* | 0.1601 | † |
| 0356-3021 | 0.0926 | 0.1003 | 0.50 | | |
| 0356-3021 | 0.0932 | 0.1010 | 0.47 | | |
| 0356-3021 | 0.0971 | 0.0998 | 0.73 | | |
| 0356-3021 | 0.0980 | 0.1008 | 0.77 | | |
| 0357-2439 | 0.0580 | 0.0580 | 1.00 | 0.0802 | 0.57 |
| 0357-2439 | 0.0592 | 0.0838 | 0.66 | 0.0802 | 0.57 |
| 0406-3105 | 0.0569 | 0.0803 | 0.002* | | |
| 0406-3105 | 0.0635 | 0.0785 | 0.05* | 0.0701 | 0.49 |
| 0406-3105 | 0.1150 | 0.0789 | 0.002* | | |
| 0412-5507 | 0.0990 | 0.0990 | 1.00 | | |
| 0412-5507 | 0.0990 | 0.0990 | 1.00 | | |
| 0412-5507 | 0.0990 | 0.0990 | 1.00 | | |
| 0422-2752 | 0.0462 | 0.0462 | 1.00 | | |
| 0422-2752 | 0.0471 | 0.0471 | 1.00 | | |
| 0422-2752 | 0.0484 | 0.0484 | 1.00 | | |
| 0424-2842 | 0.0977 | 0.1087 | 0.23* | 0.1086 | 0.33 |
| 0424-2842 | 0.1001 | 0.1114 | 0.32* | 0.1085 | 0.46 |
| 0426-2821 | 0.0940 | 0.0468 | 0.05* | | |
| 0426-2821 | 0.0950 | 0.0472 | 0.05* | | |
| 0426-2821 | 0.0952 | 0.0473 | 0.05* | | |

**Table 3.** (*continued*)

| (1) Cluster | (2) $z_{obs}$ | (3) $z_{\mathcal{L}}$ | (4) $P$ | (5) $z_{\mathcal{L}int}$ | (6) $P_{int}$ |
|---|---|---|---|---|---|
| 0427-1742 | 0.0795 | 0.0665 | 0.46 | | |
| 0427-1742 | 0.0816 | 0.0682 | 0.42 | | |
| 0427-1742 | 0.0829 | 0.0692 | 0.40 | | |
| 0430-2111 | 0.0639 | 0.1683 | 0.000* | 0.1683 | † |
| 0430-2111 | 0.0647 | 0.1673 | 0.000* | 0.1673 | † |
| 0431-3246 | 0.1162 | 0.0963 | 0.12* | | |
| 0431-3246 | 0.1179 | 0.0943 | 0.08* | | |
| 0433-2835 | 0.0433 | 0.0641 | 0.04* | 0.0579 | 0.36 |
| 0433-2835 | 0.0433 | 0.0641 | 0.04* | 0.0579 | 0.36 |
| 0434-2232 | 0.0319 | 0.0593 | 0.31* | | |
| 0434-2232 | 0.0690 | 0.0653 | 0.77 | | |
| 0438-3539 | 0.0594 | 0.0285 | 0.10* | 0.0594 | 1.00 |
| 0438-3539 | 0.0597 | 0.0286 | 0.09* | 0.0597 | 1.00 |
| 0440-3252 | 0.0434 | 0.1227 | 0.00* | | |
| 0440-3252 | 0.0799 | 0.0799 | 1.00 | | |
| 0444-2534 | 0.1150 | 0.1035 | 0.37 | | |
| 0449-5112 | 0.0914 | 0.1219 | 0.12* | 0.1016 | 0.30* |
| 0449-5112 | 0.0914 | 0.1219 | 0.12* | 0.1016 | 0.41 |
| 0449-5112 | 0.0921 | 0.1255 | 0.11* | 0.1023 | 0.43 |
| 0459-1808 | 0.0577 | 0.0994 | 0.01* | 0.1054 | 0.004* |
| 0459-1822 | 0.0427 | 0.0958 | 0.00* | | |
| 0459-1822 | 0.0430 | 0.0966 | 0.00* | | |
| 0459-1822 | 0.0797 | 0.0819 | 0.78 | | |
| 0459-1822 | 0.0800 | 0.0822 | 0.79 | | |
| 0508-3611 | 0.1168 | 0.1051 | 0.34 | 0.1155 | 0.70 |
| 0508-3611 | 0.1190 | 0.1062 | 0.31* | 0.1155 | 0.70 |
| 0512-4147 | 0.0785 | 0.1106 | 0.02* | 0.0956 | 0.32 |
| 0512-4147 | 0.0813 | 0.1080 | 0.04* | 0.0991 | 0.39 |
| 0513-4907 | 0.0564 | 0.1057 | 0.00* | | |
| 0513-4907 | 0.0913 | 0.1040 | 0.22* | 0.1016 | 0.38 |
| 0513-4907 | 0.0914 | 0.1041 | 0.22* | 0.1016 | 0.38 |
| 0513-4907 | 0.0939 | 0.1044 | 0.30* | 0.1018 | 0.47 |
| 0513-4907 | 0.0944 | 0.1050 | 0.31* | 0.1023 | 0.49 |
| 0514-3508 | 0.0996 | 0.1248 | 0.22* | 0.1220 | 0.33 |
| 0514-3508 | 0.1004 | 0.1286 | 0.20* | 0.1230 | 0.32 |
| 0515-4211 | 0.0803 | 0.1242 | 0.01* | 0.1154 | 0.36 |
| 0519-4054 | 0.0709 | 0.1052 | 0.04* | 0.0861 | 0.48 |
| 0519-4054 | 0.0770 | 0.1021 | 0.14* | 0.0854 | 0.64 |
| 0523-3131 | 0.0314 | 0.0314 | 1.00 | 0.0374 | 1.00 |
| 0523-3131 | 0.0374 | 0.0323 | 0.42 | 0.0374 | 1.00 |
| 2038-6135 | 0.0927 | 0.0539 | 0.10* | 0.0953 | 0.82 |
| 2045-6211 | 0.1058 | 0.0848 | 0.27* | | |
| 2045-6211 | 0.1100 | 0.0850 | 0.19* | 0.0975 | 0.32* |
| 2052-3610 | 0.0869 | 0.1638 | 0.19* | 0.1638 | † |
| 2056-3536 | 0.0918 | 0.1020 | 0.5534 | | |
| 2056-3536 | 0.0918 | 0.1020 | 0.5534 | | |
| 2057-4422 | 0.1426 | 0.0991 | 0.07* | 0.1426 | 1.00 |
| 2057-4422 | 0.1431 | 0.1015 | 0.06* | 0.1431 | 1.00 |
| 2102-3859 | 0.1487 | 0.1660 | 0.34 | | |
| 2102-3859 | 0.1554 | 0.1736 | 0.50 | | |
| 2106-2716 | 0.0862 | 0.0910 | 0.8196 | | |
| 2106-2716 | 0.1030 | 0.1030 | 1.0000 | | |
| 2128-4330 | 0.1047 | 0.1047 | 1.00 | | |
| 2128-4330 | 0.1047 | 0.1047 | 1.00 | | |
| 2128-4330 | 0.1050 | 0.1050 | 1.00 | | |
| 2129-2411 | 0.0626 | 0.1053 | 0.03* | | † |
| 2129-2411 | 0.0644 | 0.1052 | 0.04* | | † |
| 2129-3525 | 0.0898 | 0.1048 | 0.45 | | |
| 2129-3525 | 0.0901 | 0.1001 | 0.51 | | |
| 2130-3128 | 0.0648 | 0.0665 | 0.82 | | |
| 2130-3128 | 0.0656 | 0.0656 | 1.00 | | |
| 2135-5141 | 0.0588 | 0.0832 | 0.1470* | | |
| 2135-5141 | 0.0936 | 0.0832 | 0.5391 | | |
| 2138-3413 | 0.0764 | 0.0847 | 0.6172 | | |
| 2138-3413 | 0.0780 | 0.0865 | 0.6617 | | |
| 2139-3309 | 0.0726 | 0.0726 | 1.0000 | | |
| 2139-3309 | 0.0739 | 0.0739 | 1.0000 | | |
| 2140-3914 | 0.0658 | 0.0920 | 0.07* | 0.0799 | 0.41 |
| 2140-3914 | 0.0663 | 0.0909 | 0.08* | 0.0799 | 0.41 |
| 2142-2010 | 0.0577 | 0.1054 | 0.001* | 0.0830 | 0.46 |
| 2142-2010 | 0.0587 | 0.1044 | 0.001* | 0.0830 | 0.46 |
| 2144-4407 | 0.0605 | 0.0605 | 1.00 | | |
| 2144-4407 | 0.0639 | 0.0706 | 0.38 | 0.0706 | 0.38 |
| 2145-3251 | 0.0851 | 0.1063 | 0.3202 | | |
| 2145-3251 | 0.1066 | 0.1066 | 1.0000 | | |
| 2155-7206 | 0.0690 | 0.1427 | 0.0137 | | † |
| 2157-4324 | 0.0667 | 0.0950 | 0.07* | 0.0809 | 0.52 |
| 2157-4324 | 0.0750 | 0.1034 | 0.13* | 0.0830 | 0.62 |
| 2203-4600 | 0.0690 | 0.0985 | 0.11* | | |
| 2203-4600 | 0.0743 | 0.0984 | 0.21* | 0.0824 | 0.71 |
| 2203-4600 | 0.0782 | 0.0974 | 0.29* | 0.0824 | 0.79 |
| 2203-4600 | 0.0935 | 0.0935 | 1.00 | | |
| 2207-6546 | 0.0193 | 0.1123 | 0.000* | 0.1220 | 0.00* |
| 2207-6546 | 0.0746 | 0.1069 | 0.16* | 0.0706 | 0.77 |
| 2211-3658 | 0.0330 | 0.1595 | 0.00* | | |
| 2211-3658 | 0.0334 | 0.1563 | 0.00* | | |
| 2215-2428 | 0.0387 | 0.0530 | 0.10* | 0.0478 | 0.52 |
| 2215-2428 | 0.0388 | 0.0532 | 0.10* | 0.0478 | 0.52 |
| 2220-5528 | 0.0782 | 0.0952 | 0.26* | 0.0952 | 0.43 |
| 2220-5528 | 0.0783 | 0.0954 | 0.27* | 0.0954 | 0.43 |
| 2221-6152 | 0.0869 | 0.1253 | 0.02* | | |
| 2221-6152 | 0.1221 | 0.1221 | 1.00 | | |
| 2224-4909 | 0.0973 | 0.0918 | 0.53 | | |
| 2224-4909 | 0.0973 | 0.0918 | 0.53 | | |
| 2224-4909 | 0.1009 | 0.0910 | 0.39 | | |
| 2224-6916 | 0.0646 | 0.0680 | 0.7812 | | |
| 2224-6916 | 0.0658 | 0.0658 | 1.0000 | | |
| 2224-6916 | 0.0678 | 0.0678 | 1.0000 | | |
| 2227-4825 | 0.1019 | 0.0890 | 0.35 | | |
| 2227-4825 | 0.1025 | 0.0895 | 0.34 | | |
| 2227-4825 | 0.1035 | 0.0889 | 0.30* | | |
| 2229-2541 | 0.0336 | 0.0336 | 1.0000 | | |
| 2229-2541 | 0.0363 | 0.0363 | 1.0000 | | |
| 2233-2435 | 0.0312 | 0.0365 | 0.41 | 0.0392 | 0.40 |
| 2233-2435 | 0.0342 | 0.0342 | 1.00 | 0.0342 | 1.00 |
| 2236-1736 | 0.0727 | 0.1041 | 0.04* | 0.0837 | 0.34 |
| 2236-1736 | 0.0740 | 0.0980 | 0.06* | 0.09 | 0.37 |
| 2239-2515 | 0.0800 | 0.0712 | 0.6423 | | |
| 2240-5928 | 0.0828 | 0.1056 | 0.40 | | |
| 2240-5928 | 0.0830 | 0.1058 | 0.39 | | |
| 2241-4547 | 0.0449 | 0.0798 | 0.02* | | |
| 2241-4547 | 0.0934 | 0.0882 | 0.73 | | |
| 2242-4610 | 0.0886 | 0.1279 | 0.01* | 0.0885 | 1.00 |
| 2242-4610 | 0.0908 | 0.1312 | 0.01* | 0.1009 | 0.52 |

**Table 3.** (*continued*)

| (1) Cluster | (2) $z_{obs}$ | (3) $z_\mathcal{L}$ | (4) $P$ | (5) $z_{\mathcal{L}int}$ | (6) $P_{int}$ |
|---|---|---|---|---|---|
| 2243-1757 | 0.0707 | 0.0802 | 0.42 | | |
| 2243-1757 | 0.0731 | 0.0790 | 0.58 | | |
| 2246-5203 | 0.0950 | 0.1056 | 0.43 | 0.1056 | 0.55 |
| 2246-5203 | 0.0999 | 0.1055 | 0.70 | 0.1046 | 0.67 |
| 2246-6439 | 0.0937 | 0.0989 | 0.70 | | |
| 2246-6439 | 0.0969 | 0.0969 | 1.00 | | |
| 2256-6853 | 0.0847 | 0.1034 | 0.45 | 0.0940 | 0.53 |
| 2256-6853 | 0.0889 | 0.1012 | 0.60 | 0.0938 | 0.76 |
| 2257-6154 | 0.0843 | 0.1075 | 0.37 | 0.1075 | 0.33 |
| 2257-6154 | 0.0882 | 0.1077 | 0.44 | 0.1077 | 0.48 |
| 2309-2920 | 0.0863 | 0.1149 | 0.09* | | |
| 2309-2920 | 0.1169 | 0.1136 | 0.82 | | |
| 2314-3913 | 0.0629 | 0.1158 | 0.16* | | |
| 2314-4258 | 0.0958 | 0.1119 | 0.48 | 0.1172 | 0.40 |
| 2324-2407 | 0.0876 | 0.0973 | 0.48 | 0.0997 | 0.37 |
| 2324-2407 | 0.0887 | 0.0961 | 0.56 | 0.1009 | 0.35 |
| 2325-3639 | 0.0636 | 0.0603 | 0.8087* | 0.1038 | 0.0699* |
| 2325-3639 | 0.0939 | 0.0572 | 0.1035* | 0.1044 | 0.6615 |
| 2325-3639 | 0.0944 | 0.0575 | 0.1018* | 0.1050 | 0.6655 |
| 2329-3424 | 0.0505 | 0.0910 | 0.001* | 0.0606 | 0.44 |
| 2329-3424 | 0.0514 | 0.0928 | 0.001* | 0.0617 | 0.41 |
| 2333-3614 | 0.0960 | 0.0960 | 1.00 | | |
| 2333-3614 | 0.0964 | 0.0964 | 1.00 | | |
| 2334-3300 | 0.1095 | 0.1095 | 1.00 | | |
| 2334-3300 | 0.1109 | 0.1109 | 1.00 | | |
| 2336-4615 | 0.0671 | 0.0778 | 0.56 | | |
| 2338-2928 | 0.0624 | 0.0689 | 0.47 | | |
| 2338-2928 | 0.0628 | 0.0694 | 0.49 | | |
| 2342-2614 | 0.0513 | 0.0565 | 0.69 | 0.0616 | 0.46 |
| 2342-2614 | 0.0517 | 0.0569 | 0.69 | 0.0621 | 0.44 |
| 2353-3353 | 0.1050 | 0.0783 | 0.25* | 0.0924 | 0.37 |
| 2353-3353 | 0.1070 | 0.0797 | 0.22* | 0.0949 | 0.33 |
| 2357-6647 | 0.0730 | 0.1242 | 0.1789* | 0.1045 | 0.3579 |
| 2358-4950 | 0.0662 | 0.0662 | 1.0000 | | |
| 2358-4950 | 0.0671 | 0.0671 | 1.0000 | | |
| 2359-5418 | 0.0850 | 0.0944 | 0.6997 | | |

Notes:
*      This galaxy is rejected on the basis of the likelihood ratio test.
†      The likelihood function for this cluster is unstable. If the galaxy redshifts agree then this is adopted as the cluster redshift.

**Table 4.** Cluster redshifts obtained using the likelihood ratio test.

| (1) $\alpha(1950)$ | (2) $\delta(1950)$ | (3) $m_X$ | (4) $\mathcal{R}$ | (5) $z$ | (6) Abell |
|---|---|---|---|---|---|
| 00 01 04.10 | -51 03 33.12 | 18.536 | 56.173 | 0.118[1] | |
| 00 03 43.51 | -34 59 05.27 | 19.083 | 69.943 | 0.116[3] | A2721 |
| 00 08 58.03 | -29 08 11.40 | 17.700 | 27.547 | 0.063[2] | A2734 |
| 00 09 18.96 | -42 32 31.20 | 18.447 | 33.174 | 0.084[1] | A2736 |
| 00 11 23.50 | -43 17 15.72 | 17.582 | 27.308 | 0.122[2] | |
| 00 13 53.78 | -48 50 58.20 | 18.746 | 30.123 | 0.069[1] | |
| 00 13 56.11 | -31 36 33.12 | 18.705 | 27.515 | 0.082[2] | A2751 |
| 00 15 14.45 | -35 26 09.60 | 18.621 | 67.754 | 0.095[1] | A2755 |
| 00 16 05.71 | -42 02 58.56 | 18.169 | 27.368 | 0.092[1] | A2758 |
| 00 18 01.94 | -49 33 47.53 | 17.608 | 30.170 | 0.064[3] | A2764 |
| 00 18 04.13 | -25 58 36.48 | 19.396 | 72.301 | 0.131[3] | A0022 |
| 00 18 06.26 | -34 13 24.95 | 18.796 | 45.214 | 0.109[1] | |
| 00 20 04.99 | -53 53 36.60 | 18.612 | 35.719 | 0.098[1] | |
| 00 23 00.67 | -33 19 18.12 | 16.701 | 30.030 | 0.050[3] | S0041 |
| 00 24 19.63 | -48 49 05.52 | 18.529 | 27.640 | 0.072[2] | |
| 00 24 29.52 | -49 07 45.48 | 18.187 | 36.167 | 0.072[1] | |
| 00 25 16.90 | -30 33 15.48 | 18.547 | 34.386 | 0.119[1] | A2778 |
| 00 25 41.54 | -35 42 29.51 | 18.568 | 39.551 | 0.108[1] | |
| 00 26 07.54 | -23 53 19.32 | 19.227 | 45.402 | 0.109[3] | A0042 |
| 00 26 39.36 | -35 16 05.52 | 18.687 | 39.311 | 0.111[1] | |
| 00 26 41.47 | -30 32 44.16 | 18.262 | 29.409 | 0.103[1] | A2778 |
| 00 27 42.31 | -29 44 35.16 | 17.836 | 27.238 | 0.098[2] | A2784 |
| 00 27 53.04 | -53 41 09.60 | 18.314 | 43.219 | 0.092[1] | A2782 |
| 00 35 05.52 | -39 24 52.20 | 18.443 | 31.646 | 0.062[1] | A2799 |
| 00 35 16.63 | -31 07 06.96 | 18.675 | 41.214 | 0.062[1] | A2794 |
| 00 36 55.87 | -22 34 12.72 | 17.127 | 28.157 | 0.063[2] | A0074 |
| 00 37 26.69 | -26 25 08.40 | 18.392 | 32.296 | 0.108[2] | |
| 00 40 30.62 | -26 21 03.24 | 17.321 | 31.969 | 0.109[1] | |
| 00 40 34.94 | -28 52 22.44 | 18.589 | 40.127 | 0.107[1] | A2814 |
| 00 43 56.78 | -63 51 35.65 | 18.406 | 92.237 | 0.087[1] | A2819 |
| 00 44 31.01 | -55 00 37.44 | 17.700 | 37.740 | 0.083[1] | S0077 |
| 00 46 07.15 | -42 15 52.93 | 18.131 | 28.039 | 0.054[1] | |
| 00 48 54.96 | -28 46 19.92 | 18.503 | 37.505 | 0.051[1] | A2829 |
| 00 51 53.81 | -31 17 35.52 | 17.989 | 29.943 | 0.117[1] | |
| 00 54 03.70 | -38 09 56.87 | 18.713 | 33.606 | 0.118[1] | S0106 |
| 00 56 01.63 | -67 04 17.05 | 17.747 | 54.633 | 0.067[1] | S0112 |
| 00 56 09.38 | -34 32 20.41 | 18.835 | 30.811 | 0.104[2] | A2847 |
| 01 00 19.56 | -22 09 07.92 | 17.988 | 38.168 | 0.060[3] | A0133 |
| 01 01 49.25 | -43 07 31.44 | 17.784 | 35.768 | 0.053[1] | S0121 |
| 01 02 45.82 | -67 10 53.03 | 17.781 | 35.014 | 0.071[1] | A2864 |
| 01 07 40.08 | -46 10 28.56 | 15.983 | 31.139 | 0.023[3] | A2877 |
| 01 15 18.62 | -38 15 43.20 | 17.911 | 38.790 | 0.077[1] | A2891 |
| 01 15 44.74 | -36 50 38.76 | 18.597 | 29.208 | 0.075[1] | |
| 01 24 03.14 | -38 10 58.80 | 18.686 | 44.766 | 0.079[1] | A2911 |
| 01 31 03.41 | -27 14 19.68 | 18.069 | 29.129 | 0.084[1] | A2924 |
| 01 32 06.53 | -33 05 48.84 | 18.167 | 32.605 | 0.064[1] | S0167 |
| 01 44 01.56 | -55 39 09.71 | 18.094 | 36.684 | 0.094[1] | |
| 01 44 14.93 | -56 18 07.56 | 17.732 | 32.842 | 0.091[1] | |
| 01 52 03.19 | -35 55 08.40 | 18.765 | 43.067 | 0.123[1] | A2952 |
| 01 56 21.50 | -64 38 01.31 | 18.341 | 37.961 | 0.072[1] | S0210 |
| 01 59 43.18 | -48 29 44.88 | 18.042 | 30.151 | 0.087[1] | S0218 |
| 02 04 06.67 | -51 01 28.92 | 19.367 | 26.003 | 0.172[4] | S0222 |
| 02 11 07.32 | -47 25 09.84 | 19.334 | 64.350 | 0.115[4] | A2988 |
| 02 21 35.38 | -48 40 44.76 | 17.871 | 36.106 | 0.075[1] | A3009 |
| 02 25 51.62 | -67 14 53.16 | 18.139 | 33.617 | 0.095[1] | A3021 |
| 02 25 53.35 | -69 56 01.32 | 18.403 | 30.491 | 0.078[1] | |
| 02 27 21.19 | -34 04 22.08 | 18.722 | 34.150 | 0.075[1] | |
| 02 29 16.75 | -23 11 57.84 | 16.991 | 30.373 | 0.055[1] | |
| 02 29 51.91 | -32 15 40.32 | 18.247 | 31.220 | 0.078[1] | |
| 02 32 50.45 | -59 48 35.28 | 18.856 | 32.876 | 0.090[1] | S0280 |
| 02 34 17.28 | -19 34 56.64 | 18.692 | 47.352 | 0.088[1] | A0367 |

**Table 4.** (*continued*)

| (1) $\alpha$(1950) | (2) $\delta$(1950) | (3) $m_X$ | (4) $\mathcal{R}$ | (5) $z$ | (6) Abell |
|---|---|---|---|---|---|
| 02 42 04.42 | -26 25 55.20 | 18.688 | 33.891 | 0.135[1] | A0380 |
| 02 45 05.74 | -19 58 51.60 | 18.340 | 31.446 | 0.086[1] | |
| 02 45 31.97 | -22 50 38.76 | 18.171 | 27.184 | 0.086[2] | |
| 02 49 15.29 | -25 07 54.84 | 18.654 | 74.823 | 0.116[3] | A0389 |
| 02 49 38.28 | -71 36 41.03 | 17.293 | 27.324 | 0.069[2] | S0303 |
| 02 49 59.98 | -25 49 20.28 | 18.448 | 31.317 | 0.112[1] | A3062 |
| 02 53 06.02 | -35 37 58.08 | 17.902 | 28.754 | 0.080[1] | |
| 02 53 18.82 | -66 36 44.63 | 17.941 | 28.159 | 0.070[1] | S0311 |
| 02 58 23.45 | -36 38 10.68 | 17.527 | 29.550 | 0.092[1] | |
| 03 04 13.01 | -17 52 44.04 | 18.452 | 33.355 | 0.107[1] | A0416 |
| 03 06 03.12 | -23 52 49.44 | — | — | 0.041[3] | A0419 |
| 03 07 20.33 | -47 27 25.20 | 17.942 | 27.789 | 0.064[2] | A3093 |
| 03 09 21.96 | -27 07 36.12 | 18.319 | 50.267 | 0.068[1] | A3095 |
| 03 10 27.94 | -53 05 09.97 | 18.271 | 25.098 | 0.057[2] | |
| 03 10 45.86 | -27 21 34.20 | 17.940 | 30.697 | 0.107[2] | A3098 |
| 03 11 44.35 | -38 29 29.05 | 18.620 | 48.196 | 0.083[1] | |
| 03 13 26.95 | -29 26 30.12 | 17.442 | 26.832 | 0.067[2] | S0333 |
| 03 13 46.10 | -19 17 35.88 | 17.636 | 30.004 | 0.066[1] | A0428 |
| 03 14 30.41 | -44 06 52.56 | 17.656 | 26.648 | 0.092[2] | A3109 |
| 03 14 51.84 | -51 05 56.40 | 17.940 | 31.065 | 0.075[3] | A3110 |
| 03 15 55.32 | -44 42 22.32 | 18.689 | 43.669 | 0.076[1] | A3112 |
| 03 16 00.22 | -45 51 51.84 | 18.288 | 46.644 | 0.080[1] | A3111 |
| 03 16 09.82 | -44 25 16.68 | 18.258 | 49.031 | 0.072[3] | A3112 |
| 03 17 54.00 | -54 02 60.00 | — | — | 0.055[3] | S0339 |
| 03 20 10.54 | -45 44 56.76 | 17.172 | 33.385 | 0.070[1] | S0345 |
| 03 20 26.64 | -24 59 19.32 | 18.848 | 32.755 | 0.086[1] | |
| 03 20 50.64 | -53 20 22.20 | 17.989 | 48.744 | 0.078[1] | |
| 03 23 38.93 | -58 45 35.99 | 17.766 | 36.963 | 0.067[1] | |
| 03 25 59.02 | -53 53 06.72 | 17.017 | 39.791 | 0.059[3] | A3125 |
| 03 27 04.99 | -53 11 38.04 | 17.238 | 39.836 | 0.060[4] | |
| 03 27 23.50 | -55 52 41.88 | 18.766 | 91.874 | 0.086[3] | A3126 |
| 03 27 51.12 | -46 10 46.92 | 17.694 | 43.077 | 0.072[1] | |
| 03 29 07.94 | -52 43 04.08 | 17.264 | 59.761 | 0.059[3] | A3128 |
| 03 33 03.46 | -29 00 52.56 | 18.011 | 28.411 | 0.104[1] | |
| 03 34 04.82 | -53 50 47.40 | 17.504 | 59.296 | 0.063[1] | |
| 03 34 46.56 | -28 12 02.16 | 18.644 | 51.964 | 0.107[1] | A3141 |
| 03 35 03.77 | -39 57 15.48 | 18.795 | 55.795 | 0.103[1] | A3142 |
| 03 36 24.79 | -28 43 54.48 | 18.416 | 30.244 | 0.107[1] | |
| 03 36 28.68 | -33 19 27.48 | 18.441 | 32.114 | 0.109[1] | A3150 |
| 03 36 29.74 | -25 10 48.36 | 18.259 | 28.941 | 0.053[1] | |
| 03 36 50.30 | -40 45 08.28 | 18.400 | 25.432 | 0.062[2] | A3140 |
| 03 38 16.61 | -28 50 32.28 | 17.985 | 43.401 | 0.068[1] | A3151 |
| 03 39 05.30 | -55 13 12.00 | — | — | 0.043[3] | A3144 |
| 03 39 26.88 | -45 51 04.32 | 18.167 | 29.659 | 0.066[1] | |
| 03 41 42.29 | -53 47 57.84 | 18.082 | 62.224 | 0.058[3] | A3158 |
| 03 43 21.12 | -41 23 21.11 | 17.121 | 29.914 | 0.059[1] | S0384 |
| 03 43 38.88 | -24 25 58.44 | 19.110 | 48.669 | 0.105[3] | A0458 |
| 03 45 56.76 | -17 48 21.60 | 17.916 | 49.562 | 0.149[1] | A0462 |
| 03 46 11.57 | -18 07 35.04 | 18.384 | 32.947 | 0.039[1] | A3175 |
| 03 56 54.60 | -30 21 37.80 | 18.720 | 46.929 | 0.098[1] | A3194 |
| 03 57 47.30 | -24 39 45.36 | 18.060 | 29.039 | 0.059[1] | |
| 04 06 07.75 | -31 05 45.60 | 17.956 | 55.236 | 0.063[1] | A3223 |
| 04 12 52.37 | -55 07 40.08 | 18.621 | 32.986 | 0.099[1] | |

**Table 4.** (*continued*)

| (1) $\alpha$(1950) | (2) $\delta$(1950) | (3) $m_X$ | (4) $\mathcal{R}$ | (5) $z$ | (6) Abell |
|---|---|---|---|---|---|
| 04 22 31.39 | -27 52 12.72 | 17.006 | 29.468 | 0.048[1] | S0449 |
| 04 24 07.46 | -28 42 38.52 | 18.311 | 42.508 | 0.100[1] | S0452 |
| 04 26 10.27 | -28 21 11.52 | 17.260 | 28.386 | 0.094[1] | S0459 |
| 04 27 51.55 | -17 42 02.88 | 18.337 | 38.439 | 0.082[1] | |
| 04 30 02.90 | -21 11 16.44 | 18.867 | 31.249 | 0.064[1] | A3260 |
| 04 30 31.78 | -61 31 51.96 | 18.120 | 47.493 | 0.059[3] | A3266 |
| 04 31 07.73 | -32 46 03.01 | 17.648 | 29.637 | 0.116[1] | A3269 |
| 04 33 32.83 | -28 35 02.40 | 18.665 | 47.084 | 0.043[1] | S0471 |
| 04 34 06.53 | -22 32 39.12 | 17.305 | 36.703 | 0.069[1] | S0473 |
| 04 36 36.29 | -22 14 26.16 | 17.560 | 47.456 | 0.067[3] | A0500 |
| 04 38 25.37 | -35 39 54.00 | 18.428 | 32.734 | 0.060[1] | S0484 |
| 04 40 31.32 | -32 52 42.96 | 17.719 | 35.885 | 0.080[1] | S0491 |
| 04 44 39.60 | -25 34 33.96 | 18.778 | 49.243 | 0.115[1] | A0511 |
| 04 46 10.44 | -20 33 14.40 | 17.819 | 45.751 | 0.073[3] | A0514 |
| 04 49 21.46 | -51 12 24.85 | 18.819 | 34.581 | 0.093[1] | S0502 |
| 04 59 03.99 | -22 53 01.32 | — | — | 0.047[3] | A0533 |
| 04 59 12.67 | -18 22 15.24 | 17.540 | 50.019 | 0.080[1] | |
| 05 08 14.66 | -36 11 04.92 | 18.691 | 50.411 | 0.117[1] | A3321 |
| 05 12 10.03 | -41 47 58.93 | 18.771 | 53.283 | 0.081[1] | |
| 05 13 30.53 | -49 07 20.28 | 18.751 | 68.344 | 0.091[1] | A3330 |
| 05 14 38.47 | -35 08 29.39 | 18.879 | 32.204 | 0.100[1] | |
| 05 15 01.75 | -42 11 05.63 | 18.780 | 44.036 | 0.080[1] | A3332 |
| 05 19 48.41 | -40 54 25.21 | 18.755 | 28.825 | 0.077[1] | A3336 |
| 05 23 51.94 | -31 31 35.04 | 16.379 | 26.279 | 0.037[1] | A3341 |
| 20 35 36.36 | -61 24 36.36 | 16.742 | 43.854 | 0.071[3] | A3703 |
| 20 38 44.14 | -35 25 40.81 | 18.300 | 62.379 | 0.090[3] | A3705 |
| 20 38 53.06 | -61 35 31.20 | 17.231 | 33.611 | 0.093[1] | |
| 20 45 24.21 | -62 11 03.12 | 17.855 | 30.340 | 0.108[1] | |
| 20 48 07.92 | -52 56 04.92 | 15.741 | 44.932 | 0.047[3] | A3716 |
| 20 52 24.36 | -36 10 35.39 | 18.881 | 28.133 | 0.087[1] | |
| 20 56 37.01 | -35 36 11.52 | 18.379 | 26.185 | 0.092[2] | |
| 20 57 40.80 | -44 22 22.80 | 18.645 | 37.050 | 0.143[1] | |
| 21 02 53.91 | -38 59 51.00 | 18.843 | 40.185 | 0.149[1] | A3740 |
| 21 06 45.10 | -27 16 16.68 | 17.959 | 28.561 | 0.103[2] | S0925 |
| 21 28 19.76 | -43 30 26.64 | 18.695 | 32.332 | 0.105[1] | A3775 |
| 21 29 13.29 | -35 25 44.76 | 18.591 | 44.146 | 0.090[1] | S0952 |
| 21 29 33.96 | -24 11 38.40 | 18.446 | 29.428 | 0.063[1] | |
| 21 30 24.77 | -31 28 53.40 | 17.479 | 32.326 | 0.065[1] | |
| 21 31 03.56 | -53 51 02.52 | 17.591 | 32.325 | 0.078[3] | A3785 |
| 21 35 41.64 | -51 41 01.32 | 18.235 | 27.928 | 0.094[2] | A3796 |
| 21 38 23.71 | -34 13 37.20 | 18.132 | 28.347 | 0.077[2] | |
| 21 39 10.27 | -33 09 03.96 | 18.319 | 28.452 | 0.073[2] | |
| 21 40 30.69 | -39 14 34.80 | 17.919 | 40.173 | 0.066[1] | S0964 |
| 21 42 13.73 | -20 10 54.48 | 18.100 | 36.800 | 0.059[1] | A2372 |
| 21 44 02.14 | -44 07 21.36 | 18.068 | 29.345 | 0.062[1] | A3809 |
| 21 45 22.44 | -32 51 31.68 | 18.407 | 28.359 | 0.107[2] | A3812 |
| 21 49 31.83 | -19 48 40.32 | 19.327 | 69.692 | 0.094[3] | A2384 |
| 21 55 52.56 | -72 06 10.07 | 17.131 | 36.237 | 0.069[2] | |
| 21 57 54.51 | -43 24 12.60 | 18.610 | 33.570 | 0.070[1] | |
| 21 58 17.90 | -60 11 05.63 | 18.470 | 40.567 | 0.099[3] | A3827 |
| 22 03 57.94 | -46 00 13.68 | 18.585 | 37.280 | 0.076[1] | |
| 22 07 13.37 | -65 46 11.27 | 18.027 | 32.926 | 0.075[1] | |
| 22 15 58.87 | -24 28 39.72 | 16.674 | 28.325 | 0.039[1] | |

**Table 4.** (*continued*)

| (1)<br>$\alpha$(1950) | (2)<br>$\delta$(1950) | (3)<br>$m_X$ | (4)<br>$\mathcal{R}$ | (5)<br>$z$ | (6)<br>Abell |
|---|---|---|---|---|---|
| 22 17 02.14 | -55 28 18.84 | — | — | $0.040^3$ | A3869 |
| 22 20 41.26 | -55 28 48.00 | 18.301 | 32.149 | $0.078^1$ | |
| 22 21 00.69 | -61 52 34.68 | 18.543 | 42.807 | $0.122^1$ | |
| 22 21 29.66 | -64 30 37.44 | 18.967 | 39.439 | $0.094^3$ | S1022 |
| 22 24 14.43 | -69 16 24.24 | 17.435 | 28.356 | $0.068^2$ | A3879 |
| 22 24 22.03 | -49 09 19.08 | 18.668 | 53.851 | $0.097^1$ | A3877 |
| 22 27 23.95 | -48 25 40.43 | 18.414 | 29.884 | $0.103^1$ | A3883 |
| 22 29 31.46 | -25 41 55.68 | 16.542 | 26.256 | $0.035^2$ | |
| 22 30 05.18 | -55 03 49.33 | 18.779 | 41.799 | $0.075^3$ | A3886 |
| 22 33 36.05 | -24 35 44.88 | 17.121 | 27.097 | $0.034^1$ | A3893 |
| 22 36 35.14 | -17 36 24.48 | 17.844 | 39.867 | $0.074^1$ | A2462 |
| 22 39 49.18 | -25 15 25.20 | 17.850 | 37.495 | $0.080^2$ | |
| 22 40 39.43 | -59 28 51.95 | 18.807 | 34.216 | $0.083^1$ | |
| 22 41 12.48 | -45 47 34.80 | 18.490 | 30.986 | $0.093^1$ | |
| 22 42 59.11 | -46 10 34.32 | 18.851 | 33.839 | $0.091^1$ | A3910 |
| 22 43 29.21 | -17 57 15.84 | 17.939 | 32.379 | $0.071^1$ | A2480 |
| 22 46 16.07 | -64 39 17.28 | 18.315 | 53.466 | $0.096^1$ | A3921 |
| 22 46 44.91 | -52 03 51.84 | 18.800 | 57.446 | $0.098^1$ | A3922 |
| 22 56 55.00 | -68 53 03.48 | 18.867 | 39.081 | $0.087^1$ | S1078 |
| 22 57 20.76 | -61 54 37.44 | 17.817 | 37.884 | $0.086^1$ | |
| 22 59 33.70 | -22 17 02.76 | 19.338 | 50.752 | $0.136^3$ | A2521 |
| 23 02 54.94 | -21 38 42.36 | 18.295 | 35.898 | $0.095^3$ | A2528 |
| 23 05 55.22 | -20 09 28.44 | 18.780 | 51.898 | $0.083^3$ | A2538 |
| 23 09 08.51 | -29 20 08.16 | 18.619 | 43.000 | $0.117^1$ | |
| 23 09 36.52 | -21 50 16.80 | 18.809 | 61.707 | $0.086^3$ | A2556 |
| 23 14 36.24 | -42 58 00.48 | 17.450 | 29.636 | $0.096^1$ | S1106 |
| 23 24 09.07 | -24 07 30.00 | 18.467 | 43.173 | $0.088^1$ | A2599 |
| 23 25 31.71 | -36 39 48.61 | 17.953 | 26.387 | $0.093^2$ | |
| 23 29 08.35 | -34 24 50.03 | 17.043 | 28.615 | $0.051^1$ | A4012 |
| 23 33 32.54 | -36 14 27.23 | 18.147 | 28.694 | $0.096^1$ | |
| 23 34 15.02 | -33 00 31.68 | 18.464 | 32.987 | $0.111^1$ | |
| 23 36 58.94 | -46 15 46.07 | 18.165 | 31.477 | $0.067^1$ | S1140 |
| 23 38 58.65 | -29 28 03.00 | 17.605 | 27.438 | $0.063^1$ | |
| 23 42 37.97 | -26 14 49.56 | 17.037 | 33.693 | $0.052^1$ | A2660 |
| 23 53 05.28 | -33 53 13.92 | 17.934 | 29.970 | $0.107^1$ | S1161 |
| 23 56 20.59 | -60 55 55.20 | 19.318 | 47.358 | $0.096^3$ | A4067 |
| 23 57 53.38 | -66 47 34.44 | 18.096 | 27.320 | $0.073^2$ | S1166 |
| 23 58 08.74 | -49 50 03.48 | 17.900 | 27.576 | $0.067^2$ | |
| 23 59 44.06 | -54 18 59.04 | 18.013 | 34.972 | $0.085^2$ | |

Notes to column (7) refer to the source of the redshift:
1         This work (AAT long-slit observation)
2         This work (A.N.U. 2.3m long-slit observation)
3         Data taken from Andernach (1989)
4         Obtained by cross-referencing with Huchra (1990)